\documentclass[pra,twocolumn,floatfix,superscriptaddress,notitlepage,longbibliography]{revtex4-1}

\usepackage[normalem]{ulem}
\usepackage{graphicx, color, graphpap}% Include figure files
\usepackage{enumitem}
\usepackage{amssymb}
\usepackage{amsthm}
\usepackage{multirow}
\usepackage[colorlinks=true,citecolor=blue,linkcolor=magenta]{hyperref}
\usepackage[T1]{fontenc}
\usepackage{verbatim}
\usepackage{mathtools}
\usepackage{titlesec}
\usepackage{float}
\usepackage[linesnumbered,ruled,vlined]{algorithm2e}
\SetKwInput{kwInit}{Init}

\long\def\ca#1\cb{} %Use for commenting out: \ca...\cb

\newcommand{\ket}[1]{|#1\rangle}               %ket
\newcommand{\bra}[1]{\langle #1|}              %bra
\newcommand{\dya}[1]{\ket{#1}\!\bra{#1}}
\newcommand{\CC}{\mathcal{C}}
\newcommand{\DC}{\mathcal{D}}

\newcommand{\IC}{\mathcal{I}}

\newcommand{\OC}{\mathcal{O}}

\newcommand{\SC}{\mathcal{S}}

\newcommand{\UC}{\mathcal{U}}

\newcommand{\Tr}{{\rm Tr}}

\newcommand{\Var}{{\rm Var}}

\renewcommand{\geq}{\geqslant}
\renewcommand{\leq}{\leqslant}

\DeclareMathOperator*{\argmin}{arg\,min}
\renewcommand{\vec}[1]{\boldsymbol{#1}}  % Bold vectors instead of arrow vectors

\newcommand{\ad}{^\dagger}
\newcommand*{\id}{\openone}

\newcommand{\thv}{\vec{\theta}}
\newcommand{\gamv}{\vec{\gamma}}
\newcommand{\kvec}{\vec{k}}

{}%         Body font
{}%         Indent amount (empty = no indent, \parindent 
\begin{document}

\title{A semi-agnostic ansatz with variable structure for Variational Quantum Algorithms}

\author{M. Bilkis}
\affiliation{Fisica Teorica: Informacio i Fenomens Quantics, Departament de Fisica,
Universitat Autonoma de Barcelona, ES-08193 Bellaterra (Barcelona), Spain}
\affiliation{Theoretical Division, Los Alamos National Laboratory, Los Alamos, NM 87545, USA}

\author{M. Cerezo}
\affiliation{Theoretical Division, Los Alamos National Laboratory, Los Alamos, NM 87545, USA}
\affiliation{Center for Nonlinear Studies, Los Alamos National Laboratory, Los Alamos, NM, USA
}

\author{Guillaume Verdon}
\affiliation{Sandbox@Alphabet, Mountain View, CA, USA}
\affiliation{Institute for Quantum Computing, University of Waterloo, ON, Canada}

\author{Patrick J. Coles}
\affiliation{Theoretical Division, Los Alamos National Laboratory, Los Alamos, NM 87545, USA}

\author{Lukasz Cincio}
\address{Theoretical Division, Los Alamos National Laboratory, Los Alamos, NM 87545, USA}

\begin{abstract}
Quantum machine learning--- and specifically Variational Quantum Algorithms (VQAs)--- offers a powerful, flexible paradigm for programming near-term quantum computers, with applications in chemistry, metrology, materials science, data science, and mathematics. Here, one trains an ansatz, in the form of a parameterized quantum circuit, to accomplish a task of interest. However, challenges have recently emerged suggesting that deep ansatzes are difficult to train, due to flat training landscapes caused by randomness or by hardware noise. This motivates our work, where we present a variable structure approach to build ansatzes for VQAs. Our approach, called VAns (Variable Ansatz), applies a set of rules to both grow and (crucially) remove quantum gates in an informed manner during the optimization. Consequently, VAns is ideally suited to mitigate trainability and noise-related issues by keeping the ansatz shallow. We employ VAns in the variational quantum eigensolver for condensed matter and quantum chemistry applications, in the quantum autoencoder for data compression and in unitary compilation problems showing successful results in all cases.

\end{abstract}

\maketitle

Quantum computing holds the promise of providing solutions to many classically intractable problems. The availability of Noisy Intermediate-Scale Quantum (NISQ) devices~\cite{preskill2018quantum} has raised the question of whether these devices will themselves deliver on such a promise, or whether they will simply be a stepping stone to fault-tolerant architectures. 

Parameterized quantum circuits have emerged as one of the best hopes to make use of NISQ devices. Variational quantum algorithms (VQAs)~\cite{peruzzo2014variational,cerezo2020variationalreview,bharti2021noisy} train such circuits to minimize a cost function and consequently accomplish a task of interest. Examples of such tasks are finding ground-states~\cite{peruzzo2014variational}, solving linear systems of equations~\cite{bravo2020variational,huang2019near,xu2019variational}, simulating dynamics~\cite{yuan2019theory,cirstoiu2020variational,commeau2020variational,gibbs2021long}, factoring~\cite{anschuetz2019variational}, compiling~\cite{khatri2019quantum,sharma2019noise}, enhancing quantum metrology~\cite{beckey2020variational,koczor2020variational}, and analyzing principle components~\cite{larose2019variational,cerezo2020variational}. More generally, one may employ multiple input states to train the parameterized quantum circuit, and many data science applications have been envisioned for VQAs~\cite{biamonte2017quantum,schuld2014quest,abbas2020power,verdon2019quantum}.

Despite recent relatively large-scale implementations of VQAs~\cite{harrigan2021quantum,arute2020hartree,ollitrault2020quantum}, there are still several issues that need to be addressed to ensure that VQAs can provide a quantum advantage on NISQ devices. One issue is trainability. For instance, it has been shown that several VQA architectures become untrainable for large problem sizes due to the existence of the so-called barren plateau phenomenon~\cite{mcclean2018barren,cerezo2020cost,pesah2020absence,holmes2020barren,zhao2021analyzing,thanasilp2021subtleties}, which can be linked to circuits having large expressibility~\cite{holmes2021connecting} or generating large quantities of entanglement~\cite{sharma2020trainability,patti2020entanglement,marrero2020entanglement}. The exponential scaling caused by such barren plateaus cannot simply be escaped by changing the optimizer~\cite{cerezo2020impact,arrasmith2020effect}. However, some promising strategies have been proposed to mitigate barren plateaus, such as correlating parameters~\cite{volkoff2021large}, layerwise training~\cite{skolik2020layerwise}, and clever parameter initialization~\cite{grant2019initialization,verdon2019learning}.

The other major issue is quantum hardware noise, which accumulates with the circuit depth~\cite{wang2020noise,franca2020limitations}. This of course reduces the accuracy of observable estimation, e.g., when trying to estimate a ground state energy. However, it also leads to a more subtle and detrimental issue known as noise-induced barren plateaus~\cite{wang2020noise}.  Here, the noise corrupts the states in the quantum circuit and the cost function exponentially concentrates around its mean value. Similar to other barren plateaus, this phenomenon leads to an exponentially large precision being required to train the parameters. Currently, no strategies have been proposed to deal with noise-induced barren plateaus. Hence developing such strategies is a crucial research direction.

Circuit depth is clearly a key parameter for both of these issues. It is, therefore, essential to construct ansatzes that maintain a shallow depth to mitigate noise and trainability issues, but also that have enough expressibility to contain the problem solution. Two different strategies for ansatzes can be distinguished: either using a fixed~\cite{kandala2017hardware,cao2019quantum,bartlett2007coupled,farhi2014quantum,hadfield2019quantum} or a variable structure~\cite{grimsley2019adaptive,tang2019qubit,zhang2021mutual, rattew2019domain,chivilikhin2020mog, cincio2021machine, cincio2018learning,du2020quantum,zhang2020differentiable}. While the former is the traditional approach, the latter has recently gained considerable attention due to its versatility to address the aforementioned challenges. In variable structure ansatzes, the overall strategy consists of employing a machine learning protocol to iteratively grow the quantum circuit by placing gates that empirically lower the cost function. While these approaches address the expressibility issue by exploring specific regions of the ansatz architecture hyperspace, their depth can still grow and lead to noise-induced issues, and they can still have trainability issues from accumulating a large number of trainable parameters.

\begin{figure}[t]
\centering
\includegraphics[width=.99\columnwidth]{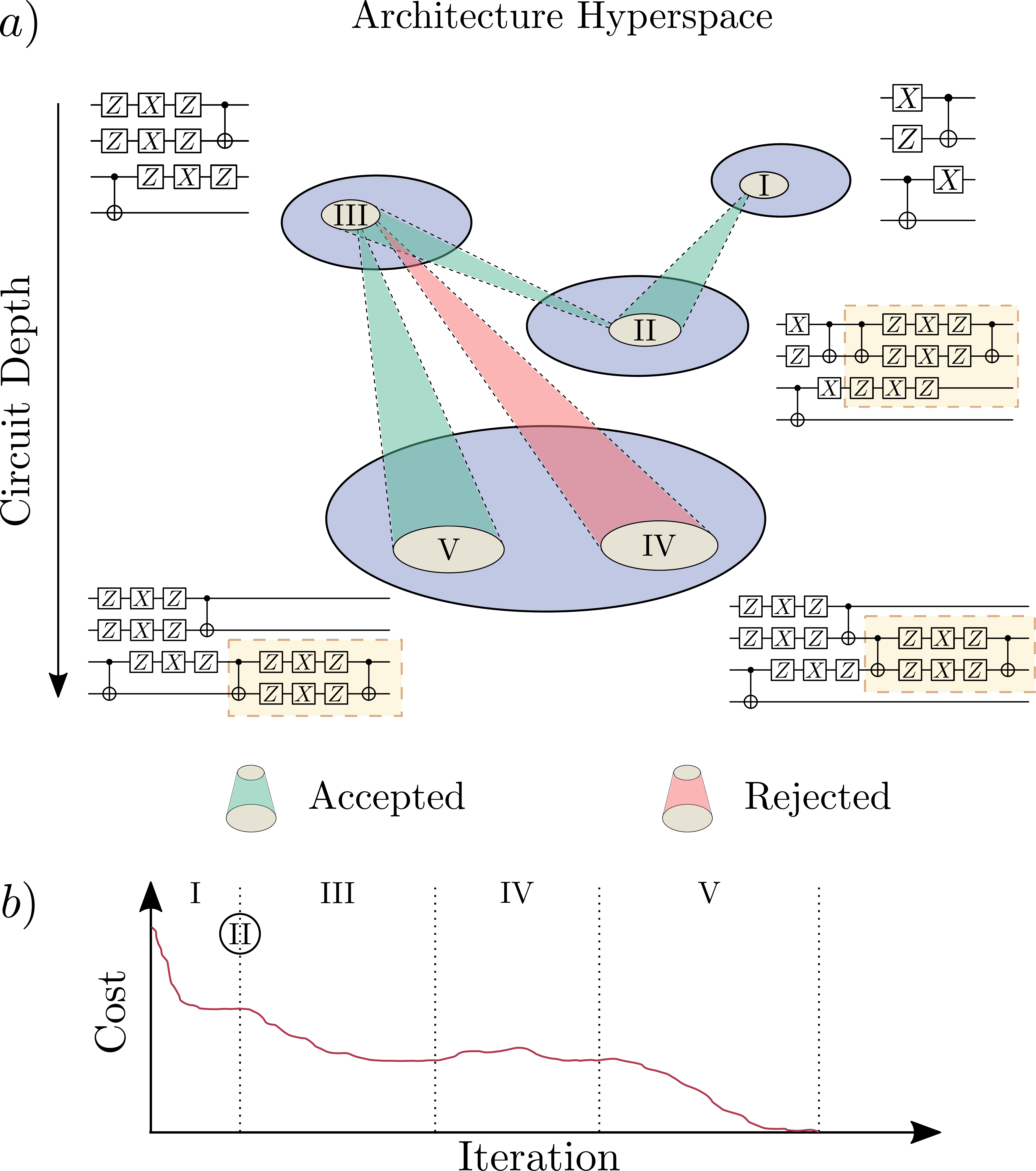}
\caption{\textbf{Schematic diagram of the VAns algorithm}. a) VAns explores the hyperspace of architectures of parametrized quantum circuits to create short depth ansatzes for VQA applications. VAns takes a (potentially non-trivial) initial circuit (step I) and optimizes its parameters until convergence. At each step, VAns inserts blocks of gates into the circuit which are initialized to the identity (indicated in a box in the figure), so that the ansatzes at contiguous steps belong to an equivalence class of circuits leading to the same cost value (step II). VAns then employs a classical algorithm to simplify the circuit by eliminating gates and finding the shortest circuit (step II to III). The ovals represent subspaces of the architecture hyperspace connected through VAns. While some regions may be smoothly connected by placing identity resolutions, VAns can also explore regions that are not smoothly connected via a gate-simplification process. VAns can either reject (step IV) or accept (step V) modifications in the circuit structure. Here $Z$ ($X$) indicates a rotation about the $z$ ($x$) axis. b) Schematic representation of the cost function value versus the number of iterations for a typical VAns implementation which follows the steps in a).}
\label{fig:schematic}
\end{figure}

In this work, we combine several features of recently proposed methods to introduce the Variable Ansatz (VAns) algorithm to generate variable structure ansatzes for generic VQA applications. As shown in Fig.~\ref{fig:schematic}, VAns iteratively grows the parameterized quantum circuit by adding blocks of gates initialized to the identity, but also prevents the circuit from over-growing by removing gates and compressing the circuit at each iteration. In this sense, VAns produces shallow circuits that are more resilient to noise, and that have less trainable parameters to avoid trainability issues. Our approach provides a simple yet effective way to address the ansatz design problem, without resorting to resource-expensive computations.

This article is fix-structured as follows. In Sec.~\ref{sec:rw} we provide background on VQA and barren plateaus, and we present a comprehensive literature review of recent variable ansatz design. We then turn to Sec.~\ref{sec:VAns}, where we present the VAns algorithm. In Sec.~\ref{sec:results} we present numerical results where we employ VAns to obtain the ground state of condensed matter and quantum chemistry systems. Here we also show how VAns can be used to build ansatzes for quantum autoencoding applications, a paradigmatic VQA implementation, and demonstrate how VAns can be applied to compile quantum circuits. Moreover, we study the noise resilience of VAns by benchmarking ground-state preparation tasks in the presence of noise. Finally, in Sec.~\ref{sec:chau} we discuss our results and present potential future research directions employing the VAns algorithm.

\section{Background}\label{sec:rw}

\subsection{Theoretical Framework}

In this work we consider generic Variational Quantum Algorithm (VQA) tasks where the goal is to solve an optimization problem encoded into a cost function of the form
\begin{equation}\label{eq:cost}
    C(\kvec,\vec{\theta})=\sum_i f_i\left(\Tr[O_i U(\kvec,\thv)\rho_i U\ad (\kvec,\thv)]\right)  \,.
\end{equation}
Here, $\{\rho_i\}$ are $n$-qubit states forming a training set, and $U(\kvec,\thv)$ is a quantum circuit parametrized by continuous parameters $\thv$ (e.g., rotation angles) and by discrete parameters $\kvec$ (e.g., gate placements). Moreover, $O_i$ are observables and $f_i$ are functions that encode the optimization task at hand. For instance, when employing the Variational Quantum Eigensolver (VQE) algorithm we have $f_i(x)=x$ and the cost function reduces to $C(\kvec,\vec{\theta})=\Tr[H U(\kvec,\thv)\rho U\ad (\kvec,\thv)]$,  where $\rho$ is the input state (and the only state in the training set) and $H$ is the Hamiltonian whose ground state one seeks to prepare. Alternatively, in a binary classification problem where the training set  is of the form $\{\rho_i,y_i\}$, with $y_i\in \{0,1\}$ being the true label, the choice $f_i(x)=(x-y_i)^2$ leads to the least square error cost. 

Given the cost function, a quantum computer is employed to estimate each term in~\eqref{eq:cost}, while the power of classical optimization algorithms is leveraged to solve the optimization task
\begin{equation}\label{eq:optimization}
    \argmin_{\kvec,\vec{\theta}} C(\kvec,\vec{\theta})\,.
\end{equation}
The success of a VQA algorithm in solving~\eqref{eq:optimization} hinges on several factors. First, the classical optimizer must be able to  efficiently train the parameters, and in the past few years, there has been a tremendous effort in developing quantum-aware optimizers~\cite{verdon2018universal,kubler2020adaptive,arrasmith2020operator,stokes2020quantum,koczor2019quantum,nakanishi2020sequential,fontana2020optimizing}. Moreover, while several choices of observables $\{O_i\}$ and functions $\{f_i\}$ can lead to different faithful cost functions (i.e., cost functions whose global optima correspond to the solution of the problem), it has been shown that global cost functions can lead to barren plateaus and trainability issues for large problem sizes~\cite{cerezo2020cost,sharma2020trainability}. Here we recall that global cost functions are defined as ones where $O_i$ acts non-trivially on all $n$ qubits. Finally, as discussed in the next section, the choice of ansatz for $U(\kvec,\thv)$ also plays a crucial role in determining the success of the VQA scheme.

\subsection{Barren Plateaus}\label{sec:BPs}

The barren plateau phenomenon has recently received considerable attention as one of the main challenges to overcome for VQA architectures to outperform their classical counterparts. Barren plateaus were first identified in~\cite{mcclean2018barren}, where it was shown that deep random parametrized quantum circuits that approximate $2$-designs have gradients that are (in average) exponentially vanishing with the system size. That is, one finds that 
\begin{equation}\label{eq:BP}
    \Var\left[\frac{\partial C(\kvec,\vec{\theta})}{\partial \theta}\right]\leq F(n)\,, \quad \text{with} \quad F(n)\in \OC\left(\frac{1}{2^n}\right)\,,
\end{equation}
where $\theta\in\thv$. From Chebyshev's inequality we have that $\Var\left[\frac{\partial C(\kvec,\vec{\theta})}{\partial \theta}\right]$ bounds the probability that the cost-function partial derivative deviates from its mean value (of zero) as
\begin{equation}\label{eq:Chebyshev}
    \Pr\left[\left|\frac{\partial C(\kvec,\vec{\theta})}{\partial \theta}\right|\geq c\right]\leq\frac{\Var[\frac{\partial C(\kvec,\vec{\theta})}{\partial \theta}]}{c^2} \,,
\end{equation}
for any $c>0$. Hence, when the cost exhibits a barren plateau, an exponentially large precision is needed to determine a cost minimizing direction and navigate the flat landscape~\cite{cerezo2020impact,arrasmith2020effect}.

This phenomenon was generalized in~\cite{cerezo2020cost} to shallow circuits, and it was shown that the locality of the operators $O_i$ in~\eqref{eq:cost} play a key role in leading to barren plateaus. Barren plateaus were later analyzed and extended to the context of dissipative~\cite{sharma2020trainability} and convolutional quantum neural networks~\cite{pesah2020absence,zhao2021analyzing}, and to the problem of learning scramblers~\cite{holmes2020barren}. A key aspect here is that circuits with large expressibility~\cite{holmes2021connecting,larocca2021diagnosing} (i.e., which sample large regions of the unitary group~\cite{sukin2019expressibility})  and which generate large amounts of entanglement~\cite{sharma2020trainability,patti2020entanglement,marrero2020entanglement} will generally suffer from barren plateaus. While several strategies have been developed to mitigate the randomness or entanglement in ansatzes prone to barren plateaus~\cite{verdon2019learning,volkoff2021large,skolik2020layerwise,grant2019initialization,pesah2020absence,zhang2020toward,bharti2020quantum,cerezo2020variational}, it is widely accepted that designing smart ansatzes which prevent altogether barren plateaus is one of the most promising applications. 

Here we remark that there exists a second method leading to barren plateaus which can even affect smart ansatzes with no randomness or entanglement-induced barren plateaus. As shown in~\cite{wang2020noise}, the presence of certain noise models acting throughout the circuit maps the input state toward the fixed point of the noise model (i.e., the maximally mixed state)~\cite{wang2020noise,franca2020limitations}, which effectively implies that the cost function value concentrates exponentially around its average as the circuit depth increases. Explicitly, in a noise-induced barren plateau one now finds that 
\begin{equation}\label{eq:NIBP}
    \left|\frac{\partial C(\kvec,\vec{\theta})}{\partial \theta}\right|\leq g(n)\,, \quad \text{with} \quad g(l)\in \OC\left(\frac{1}{q^l}\right)\,,
\end{equation}
where $q>1$ is a noise parameter and $l$ the number of layers. From Eq.~\eqref{eq:NIBP} we see that noise-induced barren plateaus will be critical for circuits whose depth scales (at least linearly) with the number of qubits.  It is worth remarking that~\eqref{eq:NIBP} is no longer probabilistic as the whole landscape flattens. Finally, we note that strategies aimed at reducing the randomness of the circuit cannot generally prevent the cost from having a noise-induced barren plateau, since here reducing the circuit noise (improving the quantum hardware) and employing shallow circuits seem to be the only viable and promising strategies to prevent these barren plateaus.

\subsection{Ansatz for Parametrized Quantum Circuits}

Here we analyze different ansatzes strategies for parametrized quantum circuits and how they can be affected by barren plateaus. Without loss of generality, a parametrized quantum circuit $U(\kvec,\thv)$ can always be expressed as
\begin{equation}\label{eq:PQC}
    U(\kvec,\thv)=\prod_j U_{k_j}(\theta_j) W_{k_j}\,,
\end{equation}
where $W_{k_j}$ are fixed gates, and where $U_{k_j}(\theta_j)=e^{-i \theta_j G_{k_j}}$ are unitaries generated by a Hermitian operator $G_{k_j}$ and parametrized by a continuous parameter $\theta_j\in\thv$. In a \textit{fixed structure} ansatz, the discrete parameters $k_j\in\kvec$ usually determine the type of gate, while in a \textit{variable structure} ansatz they can also control the gate placement in the circuit. 

\subsubsection{Fixed Structure Ansatz}

Let us first discuss fixed structure ansatzes. A common architecture with fixed structure is the layered Hardware Efficient Ansatz (HEA)~\cite{kandala2017hardware}, where the gates are arranged in a brick-like fashion and act on alternating pairs of qubits. One of the main advantages of this ansatz is that it employs gates native to the specific device used, hence avoiding unnecessary depth overhead arising from compiling non-native unitaries into native gates. This type of ansatz is \textit{problem-agnostic}, in the sense that it is expressible enough so that it can be generically employed for any task. However, its high expressibility~\cite{holmes2021connecting} can lead to trainability and barren plateau issues.

As previously mentioned, designing smart ansatzes can help in preventing barren plateaus. One such strategy are the so-called \textit{problem inspired} ansatzes. Here the goal is to encode information of the problem into the architecture of the ansatz so that the optimal solution of~\eqref{eq:optimization} exists within the parameter space without requiring high expressibility. Examples of these fixed structure ansatzes are the Unitary Coupled Cluster (UCC) Ansatz~\cite{cao2019quantum,bartlett2007coupled} for quantum chemistry and the Quantum Alternating Operator Ansatz (QAOA) for optimization~\cite{farhi2014quantum,hadfield2019quantum}. However, while these ansatzes might not exhibit expressibility-induced barren plateaus, they usually require deep circuits to be implemented, and hence are very prone to be affected by noise-induced barren plateaus~\cite{wang2020noise}. 

\subsubsection{Variable Structure Ansatzes}

To avoid some of the limitations of these fixed structure ansatzes, there has recently been great effort put forward towards  developing variable ansatz strategies for parametrized quantum circuits~\cite{grimsley2019adaptive,tang2019qubit,zhang2021mutual, rattew2019domain,chivilikhin2020mog, cincio2021machine, cincio2018learning,du2020quantum,zhang2020differentiable}. Here, the overall strategy consists of iteratively changing the quantum circuit by placing (or removing) gates that empirically lower the cost function. In what follows we briefly review some of these variable ansatz proposals. 

The first proposal for variable ansatzes for quantum chemistry was introduced in~\cite{grimsley2019adaptive} under the name of ADAPT-VQE. Here, the authors follow a circuit structure similar to that used in the UCC ansatz and propose to iteratively grow the circuit by appending gates that implement fermionic operators chosen from a pool of single and double excitation operators. At each iteration, one decides which operator in the pool is to be appended, which can lead to a considerable overhead if the number of operators in the pool is large. Similarly to fix structure UCC ansatzes, the mapping from fermions to qubits can lead to prohibitively deep circuits. This issue can be overcome using the qubit-ADAPT-VQE~\cite{tang2019qubit} algorithm, where the pool of operators is modified in such a way that only easily implementable gates are considered. However, the size of the pool still grows with the number of qubits. In this context, trainability issues have been tackled in~\cite{sim2021adaptive}, where the parameter optimization is turned into a series of optimizations each performed on a subset of the whole parameter set; an heuristic method is proposed in order to remove gates and add new ones, which in turn allowed the authors to optimize circuits naively containing more than a thousand parameters. In addition, Ref.~\cite{zhang2021mutual} studies how the pool of operators can be further reduced by computing the mutual information between the qubits in classically approximated ground state. We remark that estimations of the mutual information have also been recently employed to reduce the depth of fixed structured ansatzes~\cite{tkachenko2020correlation}. We refer the reader to~\cite{claudino2020benchmarking} for a detailed comparison between ADAPT-VQE and UCC ansatzes. 
Despite constituting a promising approach, it is unclear whether ADAPT-VQE and its variants will be able to overcome typical noise-induced trainability problems as the systems under study are increased in size. Moreover, due to its specific quantum chemistry scope, the application of these schemes is limited. 

A different approach to variable ansatzes that has gained considerable attention are machine-learning-aided evolutionary algorithms  that upgrade individuals (quantum circuits) from a population. Noticeably, the presence of quantum correlations makes it so that it is not straightforward to combine features between circuits during the evolution, as simply merging two promising circuits does not necessarily lead to low cost function values. Thus, only random mutations have been considered so far. An example of this method is found in the 
Evolutionary VQE (EVQE)~\cite{rattew2019domain}, where one smoothly explores the Hilbert space by growing the circuit with identity-initialized blocks of gates and randomly removing sequences of gates. As suggested by the authors, this might avoid entering regions leading to barren plateaus.  Another example of an evolutionary algorithm is the Multi-objective Genetic VQE (MoG-VQE)~\cite{chivilikhin2020mog}, where one uses  block-structured ansatz and simultaneously optimizes both the energy and number of entangling gates. Evolutionary algorithms constitute a promising approach to ansatzes design, they nevertheless come at the cost of high quantum-computational resources to evolve populations of quantum circuits.

A different machine learning approach to discover ansatz structures  has been considered in~\cite{cincio2018learning,cincio2021machine} where the goal is to obtain a short depth version of a given unitary.  Given specific quantum hardware constraints (such as connectivity, noise-model as represented by quantum channels or available gates), an algorithm grows and modifies the structure of a parametrized quantum circuit to best match the action of the trained circuit with that of the target unitary.  At each iteration, a parallelization and compression procedure is applied. This method was able to discover a short-depth version of the swap test to compute state overlaps~\cite{cincio2018learning}, and in~\cite{cincio2021machine} it was shown to drastically improve the discovered circuit performance in the presence of noise. Moreover, this technique has recently been tested in large-scale numerics for combinatorial optimization problems~\cite{liu2021layer}. In addition, in~\cite{ostaszewski2021structure} the authors present a different iterative algorithm where single-qubit rotations are used for growing the circuit, hence leading to a scheme with limited expressibility power. 

Finally, in the recent works of Refs.~\cite{du2020quantum,zhang2020differentiable,pirhooshyaran2021quantum} the authors employ tools from auto-machine learning to build variable ansatzes. Specifically, in~\cite{du2020quantum} the authors make use of the supernet and weight sharing strategies from 
neural network architecture search~\cite{elsken2019neural}, while in~\cite{zhang2020differentiable} the proposal is based on a generalization of the differentiable neural architecture search~\cite{liu2018darts} for quantum circuits. In~\cite{pirhooshyaran2021quantum} the authors study the design of quantum circuits for multi-label classification: policy gradient methods are used to train a neural network, so to decide which gates will compose candidate quantum circuits. We remark that while promising, the latter methods require quantum-computational resources which considerably grow with the problem sizes. 

In the next section, we present a task-oriented NISQ-friendly approach to the problem of ansatzes design. In the context of the literature, the approach presented here generalizes  the work in~\cite{cincio2018learning,cincio2021machine} as a method to build potentially trainable short-depth ansatz for Variation Quantum Algorithm tasks. Unlike previous methods, the algorithm introduced here not only grows the circuit but more importantly employs classical routines to remove quantum gates in an informed manner during the optimization. In addition, our method can be used in a wide class of problems within the realm of variational quantum algorithms (where the cost function is usually defined as the expectation of a Hamiltonian), but also within more general contexts (where the loss function can take more general forms such as mean-squared error, log likelihoods, etc). Finally, below we showcase the performance of VAns in the presence of hardware noise, a benchmark notably absent in many other variable ansatzes proposals.

\section{The Variable Ansatz (VAns) Algorithm}\label{sec:VAns}

\subsection{Overview}

The goal of the VAns algorithm is to adaptively construct shallow and trainable ansatzes for generic quantum machine learning applications using parametrized quantum circuits. Let us define as $\CC_l$ the architecture hyperspace of quantum circuits of depth $l$, where a single layer is defined as gates acting in parallel. VAns takes as input:
\begin{itemize}
    \item A cost function $C(\kvec,\thv)$ to minimize.
    \item A dictionary $\DC$ of parametrized gates that compile to the identity. That is, for  $V(\gamv)\in\DC$ there exists a set of parameters $\gamv^*$ such that $V(\gamv^*)=\id$.
    \item An initial circuit configuration $U^{(0)}(\kvec,\thv)\in\CC_{l_0}$ of depth $l_0$ .
    \item Circuit \texttt{Insertion} rules which stochastically take an element $V(\gamv^*)\in\DC$ and append it to the circuit. The insertion is a map $\IC:\CC_l\rightarrow \CC_{l'}$ with $l'\geq l$. 
    \item Circuit \texttt{Simplification} rules to eliminate unnecessary gates, redundant gates, or gates that do not have a large effect on the cost. The simplification is a map $\SC:\CC_l\rightarrow \CC_{l'}$ with $l'\leq l$. 
    \item An optimization algorithm for the continuous parameters $\thv$, and an optimization algorithm for the discrete parameters $\kvec$.
\end{itemize}
Given these inputs, VAns outputs a circuit architecture and set of parameters that approximately minimize~\eqref{eq:optimization}. In what follows we describe the overall structure of VAns (presented in Algorithm~\ref{alg1}), and in the next sections we provide additional details for the \texttt{Insertion} and \texttt{Simplification} modules. In all cases, the steps presented here are aimed at giving a general overview of the method and are intended to be used as building blocks for more advanced versions of VAns.

\begin{figure}[t]
\centering
\includegraphics[width=.99\columnwidth]{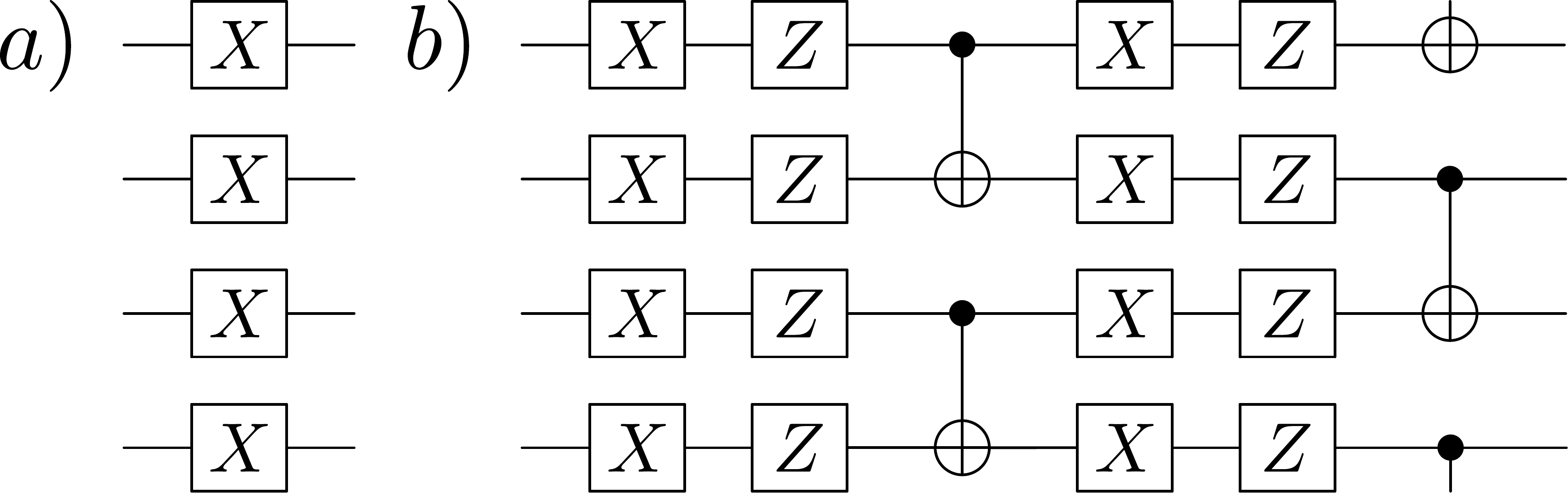}
\caption{\textbf{Examples of initial circuit configurations for VAns.} VAns take as input an initial structure for the parametrized quantum circuit. In (a)  we depict a separable product ansatz which generates no entanglement between the qubits. On the other hand, (b) shows two layers of a shallow alternating Hardware Efficient Ansatz where neighboring  qubits are initially entangled.  Here $Z$ ($X$) indicates a rotation about the $z$ ($x$) axis. }
\label{fig:initialansatz}
\end{figure}

The first ingredient of VAns (besides the cost function, which is defined by the problem at hand) is a  dictionary $\DC$ of parametrized gates that can compile to identity and  which VAns will employ to build the ansatz. A key aspect here is that $\DC$ can be composed of any set of gates, so that one can build a dictionary specifically tailored for a given application. For instance, for problems with a given symmetry,  $\DC$ can contain gates preserving said symmetry~\cite{gard2020efficient}. In addition, it is usually convenient to have the unitaries in $\DC$ expressed in terms of gates native to the specific quantum hardware employed, as this avoids compilation depth overheads.

Once the gate dictionary is set, the ansatz is initialized to a given configuration $U^{(0)}(\kvec,\thv)$. As shown in Algorithm~\ref{alg1}, one then employs an optimizer to train the continuous parameters $\thv$ in the initial ansatz until the optimization algorithm converges. In Fig.~\ref{fig:initialansatz}, we show two non-trivial initialization strategies employed in our numerical simulations (see Section~\ref{sec:results}). In Fig.~\ref{fig:initialansatz}(a) the circuit is initialized to a separable product ansatz which generates no entanglement, while in Fig.~\ref{fig:initialansatz}(b) one initializes to a shallow alternating Hardware Efficient Ansatz such that neighboring  qubits are entangled. While the choice of an appropriate initial ansatz can lead to faster convergence, VAns can in principle transform a simple initial ansatz into a more complex one as part of its architecture search. 

From this point, VAns enters a nested optimization loop. In the outer loop, VAns explores the architecture hyperspace to optimize the ansatz's discrete parameters $\kvec$ that characterize the circuit structure. Then, in the inner loop, the ansatz structure is fixed and the continuous parameters $\thv$ are optimized. 

At the start of the outer loop, VAns employs its \texttt{Insertion} rules to stochastically grow the circuit. The fact that these rules are stochastic guarantees that different runs of VAns explore distinct regions of the architecture hyperspace.  As previously mentioned, the gates added to the circuit compile to the identity so that circuits that differ by gate insertions belong to an equivalence class of circuits leading to the same cost function value. As discussed below, the \texttt{Insertion} rules can be such  that they depend on the current circuit they act upon. For instance, VAns can potentially add entangling gates to qubits that were are not previously connected via two-qubit gates. 

To prevent the circuit from constantly growing each time gates are inserted, VAns follows the \texttt{Insertion} step by a \texttt{Simplification} step. Here, the goal is to determine if the circuit depth can be reduced without significantly modifying the cost function value in a systematic way, as proposed in~\cite{maslov2008quantum}.  This is a fundamental step of VAns as it allows the algorithm to explore and jump between different regions of the architecture hyperspace which might not be trivially connected.   Moreover, unlike other variable ansatz strategies which continuously increase the circuit depth or which randomly remove gates, the \texttt{Simplification} step allows VAns to find short depth ansatzes by deleting gates in an informed manner.

Taken together, \texttt{Insertion} and \texttt{Simplification} provide a set of discrete parameters $\kvec$. However, to determine if this new circuit structure can improve the cost function value it is necessary to enter the inner optimization loop and train the continuous parameters $\thv$. When convergence in the optimization is reached, the final cost function value is compared to the cost in the previous iteration. Updates that lead to equi-cost values or to smaller costs are accepted, while updates leading to higher cost functions are accepted with exponentially decaying probability in a manner similar to a Metropolis-Hastings step~\cite{hastings1970monte}. Here one accepts an update which increases the cost value with a probability given by $\exp{( -  \beta \frac{\Delta \mathcal{C}}{\mathcal{C}_0} )}$, with $\frac{\Delta \mathcal{C}}{\mathcal{C}_0}$ being increment in the cost function with respect to the initial value, and $\beta > 0$ a  ``temperature'' factor.
The previous optimizations in inner and outer loops are repeated until a termination condition $f_\texttt{Term}$ is reached, \textit{e.g.} distance to a target cost function value (if a lower bound is available), maximum VAns iteration number, or an user-specified function that might depend on variables such as circuit structure and cost value reached.

\begin{algorithm}[t]\label{alg1}
    \DontPrintSemicolon
    \KwIn{Cost function $C(\kvec,\thv)$; initial circuit $U^{(0)}(\kvec,\thv)$; dictionary of gates $\DC$; \texttt{Insertion} rules which take gates from $\DC$ and appends them to a circuit; \texttt{Simplification} rules; optimization algorithm \texttt{Opt}$_C$ for continuous parameters $\thv$; optimization algorithm \texttt{Opt}$_D$ for discrete parameters which accepts or rejects an ansatz update given changes in the cost function value; termination condition function $f_\texttt{Term}(n, C(\kvec,\thv),U(\kvec, \thv)), \; n\in \mathbb{N}$}
    \KwOut{Optimized ansatz $U^{(f)}$.}
    \kwInit{Randomly initialize the parameters $\thv$;
    initialize the ansatz $U^{(f)}\gets U^{(0)}(\kvec,\thv)$; 
     $C^{(f)}\gets 0$; 
    $\kvec^{(f)}\gets \kvec$; 
    $\thv^{(f)}\gets \thv$;
    $\UC(\kvec,\thv)\gets \id$;
    $\texttt{Term} \leftarrow false$; $n \leftarrow 0$.
    }
    Optimize $\thv$ with \texttt{Opt}$_C$ and store result in $\thv^{(f)}$; $C^{(f)}\gets C(\kvec,\thv)$.\;
    \While{\texttt{Term} is false}{
    $n \leftarrow n + 1$\;
    $\texttt{Accept}\gets\textit{false}$\;
    \While{\texttt{Accept} is false}{
    Use \texttt{Insertion} in $U^{(f)}$ and store new sets of discrete parameters, continuous parameters and ansatz in $\thv$, $\kvec$, and  $\UC(\kvec,\thv)$, respectively.\;
    Use \texttt{Simplification} rules 1-5 on $\UC(\kvec,\thv)$ and store new sets of discrete parameters, continuous parameters and ansatz in $\thv$, $\kvec$, and  $\UC(\kvec,\thv)$, respectively.\;
    Optimize continuous parameters in $\UC(\kvec,\thv)$ with \texttt{Opt}$_C$ and store result in $\thv$; $C^{(f)}\gets C(\kvec,\thv)$.\;
    Use \texttt{Simplification} rule 6\;
    Given $C(\kvec,\thv)$ and $C^{(f)}$, optimize discrete parameters in $\UC(\kvec,\thv)$ with \texttt{Opt}$_D$ and store result in \texttt{Accept}.\;
    }
    $\kvec^{(f)}\gets \kvec$.\; 
    $\thv^{(f)}\gets \thv$.\;
    $U^{(f)}\gets \UC(\kvec,\thv)$.\;
    $C^{(f)}\gets C(\kvec,\thv)$.\;
    $\texttt{Term} \leftarrow f_{\texttt{Term}}(n,C^{(f)},U^{(f)})$
    }
    \caption{Pseudo-code for VAns}
 \end{algorithm}

\subsection{\texttt{Insertion} method}

\begin{figure}[t]
\centering
\includegraphics[width=.65\columnwidth]{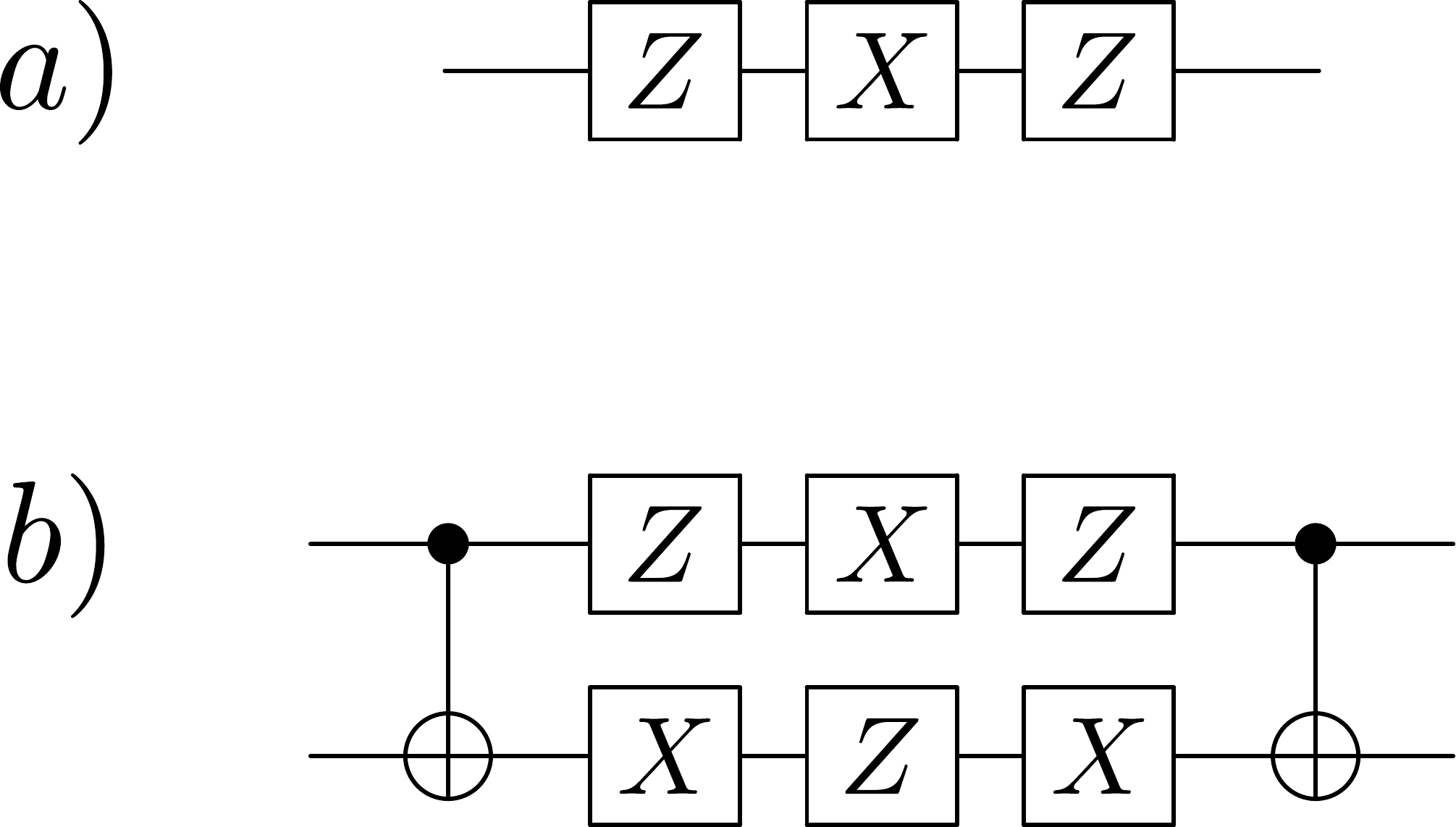}
\caption{\textbf{Circuits from the dictionary $\DC$ used during the \texttt{Insertion} steps.} Here we show two  types of the parametrized gate sequences composed of CNOTs and rotations about the $z$ and $x$ axis. Specifically, one obtains the identity if the rotation angles are set to zero. Using the circuit in (a), one inserts a general unitary acting on a given qubit, while the circuit in (b) entangles the two qubits it acts upon.}
\label{fig:blocks}
\end{figure}

As previously mentioned, the \texttt{Insertion} step stochastically grows the circuit by inserting into the circuit a parametrized block of gates from the dictionary $\DC$ which compiles to the identity. It is worth noting that in practice one can allow some deviation from the identity to reach a larger gate dictionary $\DC$. In turn, this permits VAns to explore regions of the architecture hyperspace that could otherwise take several iterations to be reached. In Fig.~\ref{fig:blocks} we show examples of two parametrized quantum circuits that can compile to the identity. 

There are many choices for how VAns determines which gates are chosen from $\DC$ at each iteration, and where they should be placed. When selecting gates, we have here taken a uniform sampling approach, where every sequence of gates in $\DC$ has an equal probability to be selected. While one could follow a similar approach for determining where said gates should be inserted, this can lead to deeper circuits with regions containing an uneven number of CNOTs. In our heuristics, VAns has a higher probability to place two-qubit gates acting on qubits that were otherwise not previously connected or shared a small number of entangling gates.

\subsection{\texttt{Simplification} method}

\begin{figure}[t]
\centering
\includegraphics[width=.85\columnwidth]{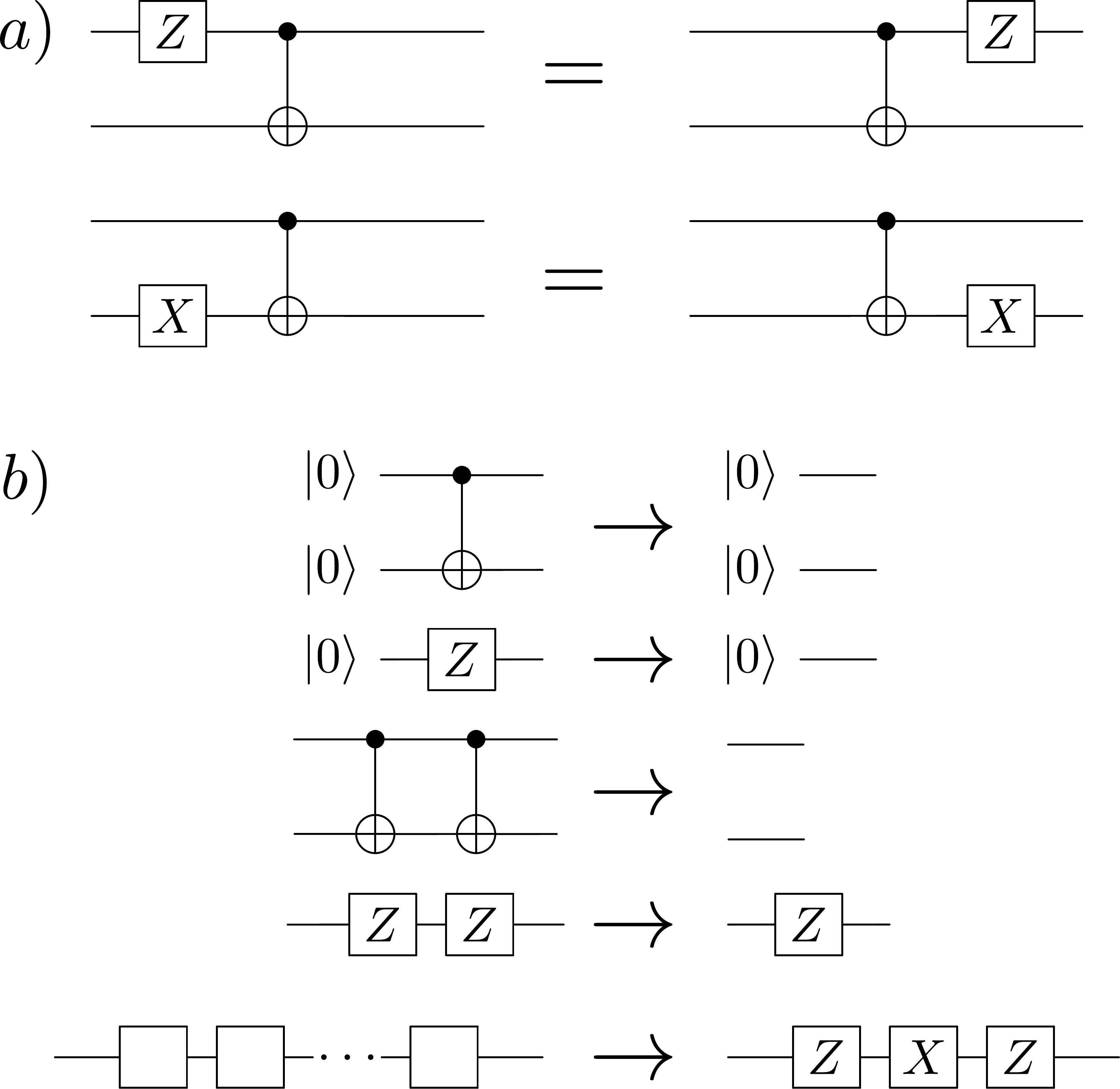}
\caption{\textbf{Rules for the  \texttt{Simplification} steps.} (a) Commutation rules used by VAns to move gates in the circuit. As shown, one can commute a CNOT with a rotation $Z$ ($X$) about the $z$ ($x$) axis acting on the control (target) qubit. (b) Example of simplification rules used by VAns to reduce the circuit depth. Here we assume that the circuit is initialized to $\ket{0}^{\otimes n}$. }
\label{fig:simplification}
\end{figure}

The \texttt{Simplification} steps in VAns are aimed at eliminating unnecessary gates, redundant gates, or gates that do not have a large effect on the cost. For this purpose, \texttt{Simplification} moves gates in the circuit using the commutation rules shown in  Fig.~\ref{fig:simplification}(a) to group single qubits rotations and CNOTs together. 
Once there are no further commutations possible,  the circuit is scanned and a sequence of simplification rules are consequently applied. For instance, assuming that the input state is initialized to $\ket{0}^{\otimes n}$, we can define the following set of  simplification rules. 
\begin{enumerate}
\item CNOT gates acting at the beginning of the circuit are removed.
\item Rotations around the $z$-axis acting at the beginning of the circuit are removed. 
\item Consecutive CNOT sharing the same control and target qubits are removed.
\item Two or more consecutive rotations around the same axis and acting on the same qubit are compiled into a single rotation (whose value is the sum of the previous values).
\item If three or more single-qubit rotations are sequentially acting on the same qubit, they are simplified into a general single-qubit rotation of the form $R_z(\theta_1) R_x(\theta_2) R_z(\theta_3)$ or $R_x(\theta_1) R_z(\theta_2) R_x(\theta_3)$ which has the same action as the previous product of rotations ---note that in the figures we denote such rotations by X or Z ---.
\item Gates whose presence in the circuit does not considerably reduce the cost are removed. 
\end{enumerate}
Rules $(1)-(5)$ are schematically shown in Fig.~\ref{fig:simplification}(b). We remark that a crucial feature of these \texttt{Simplification} rules is that they can be performed using a classical computer that analyzes the circuit structure and hence do not lead to additional quantum-computation resources.

As indicated by step $(6)$, the \texttt{Simplification} steps can also delete gates whose presence in the circuit does not considerably reduce the cost. Here, given a parametrized gate, one can remove it from the circuit and compute the ensuing cost function value. If the resulting cost is increased by more than some threshold value, the gate under consideration is removed and the simplification rules $(1)-(5)$ are again implemented. Here, one can use information from the inner optimization loop to find candidate gates for removal. For instance, when employing a gradient descent optimizer, one may attempt to remove gates whose parameters lead to small gradients. Note that, unlike the  simplification steps $(1)-(5)$ in Fig.~\ref{fig:simplification}(b), when using the deletion process in $(6)$  one needs to call a quantum computer to estimate the cost function, and hence these come at an additional quantum-resource overhead which scales linearly with the number of gates one is attempting to remove. 

\begin{figure}[t]
\centering
\includegraphics[width=.65\columnwidth]{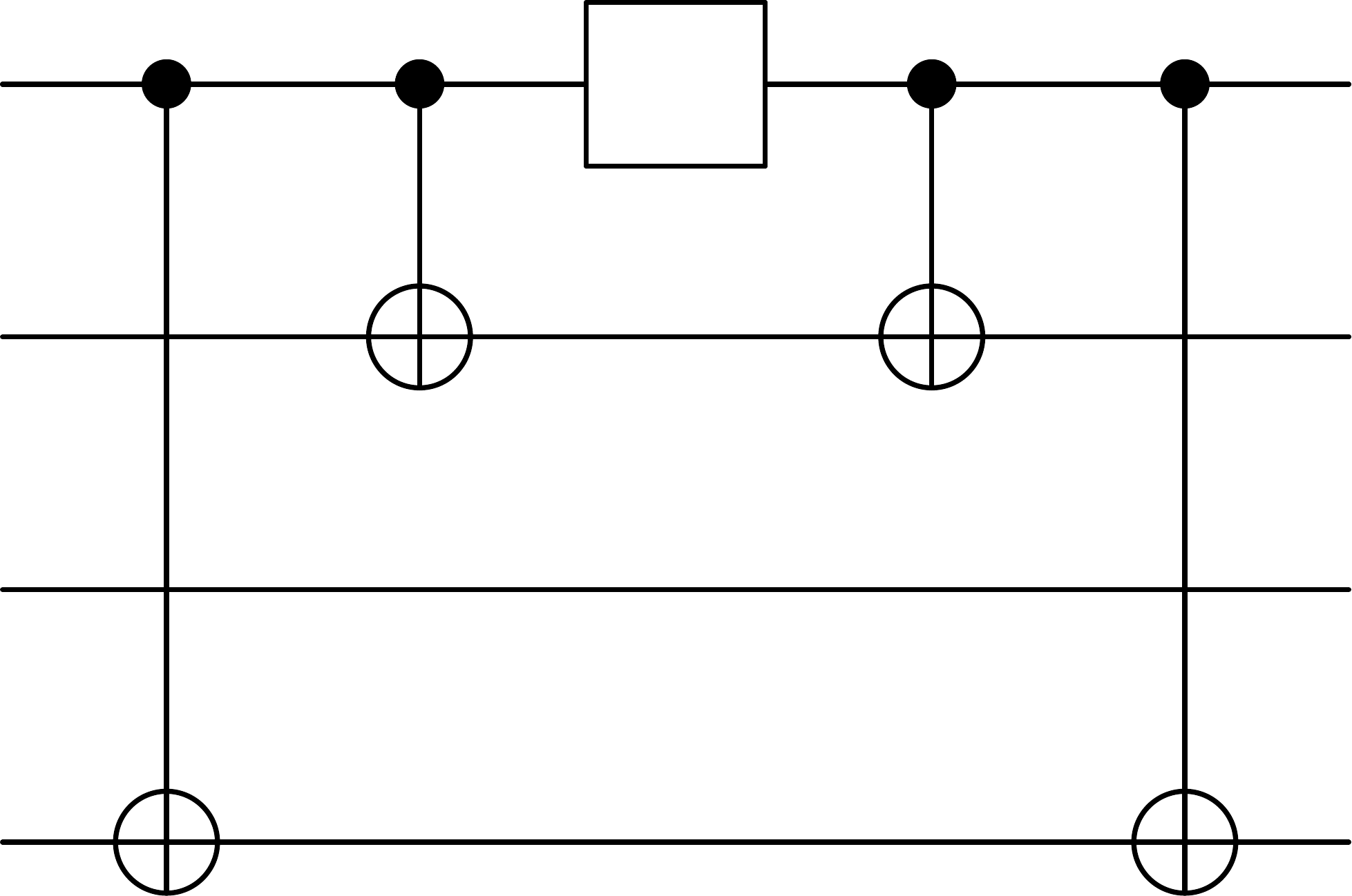}
\caption{\textbf{Circuit obtained from VAns}. Shown is a non-trivial circuit structure that can be obtained by VAns using the  \texttt{Insertion} and \texttt{Simplification} steps and the gate dictionary in Fig.~\ref{fig:blocks}.   }
\label{fig:new}
\end{figure}

An interesting aspect of the \texttt{Simplification} method is that it allows VAns to obtain circuit structures that are not contained in the initial circuit $U^{(0)}(\kvec,\thv)$ or in the gate dictionary $\DC$, and hence to explore new regions of the architecture hyperspace. For instance, using the gate dictionary in Fig.~\ref{fig:blocks},  VAns can obtain a gate structure as the one shown in Fig.~\ref{fig:new}.

\subsection{Scaling of VAns}

With the previous overview of the VAns method in mind, let us now discuss the computational complexity arising from using VAns versus that of using a standard fix architecture scheme. In the following discussion we will not include any computational cost or complexity of the optimizer as we assume that the same tools could be used for fixed or variable ansatzes.

Firstly, we note that any additional computational cost comes due to circuit manipulations, meaning that we should study the scaling of the \texttt{Insertion} and \texttt{Simplification} methods. On the one hand, adding gates via \texttt{Insertion} is stochastic, and independent of the number of qubits or the current circuit depth, that is:  its complexity is always in $\OC(1)$. Then, removing gates via \texttt{Simplification} has a cost which increases with the number of gates in the circuit. If we have $M$ gates, then running the Simplification scheme has a cost $\OC(M)$. We note that such computational complexity is similar to that of using gradient-free versus gradient-based methods, as the computational cost of the latter also scale as $\OC(M)$. Notably,  since the goal of VAns is to produce short-depth circuits, the algorithm itself tries to reduce its own computational cost during training.  As we sill see below in our numerical examples, VAns is always able to find short-depth circuits whose solutions are better than those arising from fixed structure ansatzes, meaning that the extra complexity of VAns can be justified in terms of its performance.

\subsection{Mitigating the effect of barren plateaus}

\subsubsection{General considerations}

Let us now discuss why VAns is expected to mitigate the impact of barren plateaus.

First, consider the type of barren plateaus that are caused by the circuit approaching an approximate 2-design~\cite{mcclean2018barren}. Approximating a 2-design requires a circuit that both has a significant number of parameters and also has a significant depth~\cite{brandao2016local,dankert2009exact,harrow2009random,harrow2018approximate,haferkamp2022randomquantum}. Hence, reducing either the number of parameters or the circuit depth can combat these barren plateaus. Fortunately, VAns attempts to reduce both the number of parameters and the depth. Consequently, VAns actively attempts to avoid approximating a 2-design.

Second, consider the barren plateaus that are caused by hardware noise~\cite{wang2020noise}. For such barren plateaus, it was shown that the circuit depth is the key parameter, as the gradient vanishes exponentially with the depth. As VAns actively attempts to reduce the number of CNOTs, it also reduces the circuit depth. Hence VAns will mitigate the effect of noise-induced barren plateaus by keeping the depth shallow during the optimization. Evidence of such a mechanism can be found in the next section, where we consider simulations under the presence of noise. In such a case, VAns automatically adjusts the circuit layout in such a way that the cost function reaches a minima, which translates to short-depth circuits in noisy scenarios.

\subsubsection{Specific applications}
While in the previous subsection we have presented general arguments as to why VAns can improve trainability,  here we instead present a practical method that combines VAns with the recent techniques of Ref.~\cite{sack2022avoiding} for mitigating barren plateaus using classical shadows. For completeness, we briefly review the results in~\cite{sack2022avoiding}.  Firstly, as noted in Section~\ref{sec:BPs} it is known that the presence of barren plateaus is intrinsically related to the entanglement generated in the circuit~\cite{sharma2020trainability,patti2020entanglement,marrero2020entanglement}. That is, circuits generating large amounts of entanglement are prone to barren plateaus. With this remark in mind, the authors in~\cite{sack2022avoiding} propose to detect the onset of a barren plateau by monitoring the entanglement in the state at the output of the circuit. This can be achieved by computing, via classical shadows~\cite{huang2020predicting}, the second R\'enyi entanglement entropy $S_2(\rho_R)=-\log(\Tr[\rho_R]^2)$, where $\rho_R=\Tr_{\overline{R}}[U(\kvec,\thv)\rho_i U\ad (\kvec,\thv)]$ denotes a reduced state on a subset of $R$ qubits. As such, if $S_2(\rho_R)$ approaches the maximal possible entanglement of the $S$ qubits, given by the so-called Page value $S^{\text{page}}\sim k\log(2)-\frac{1}{2^{n-2k+1}}$, one knows that the optimization is leading to a region of high entanglement, and thus of barren plateaus. The key proposal in~\cite{sack2022avoiding} is to tune the optimizer (e.g., by controlling the gradient step) so that regions of large entropies are avoided. This technique is shown to work well with an identity block initialization~\cite{grant2019initialization}, whereby the parameters in the trainable unitary are chosen such that $U(\kvec,\thv)=\id$ at the start of the algorithm. Note that, in principle, this is still a fixed ansatz method, as some circuit structure has to be fixed beforehand, and as no gates are ever removed. Hence, the methods in~\cite{sack2022avoiding} can be readily combined with VAns to variationally explore the architecture hyperspace while keeping track of the reduced state entropy. In practice, this means that one can modify the VAns update rule to allow for steps that do not significantly increase entropy, while favouring steps that keep the entropy constant, or even that reduce it (\textit{e.g.} by removing gates during the \texttt{Simplification} modules).

\section{Numerical Results}\label{sec:results}
In this section, we present heuristic results obtained from simulating VAns to solve paradigmatic problems in condensed matter, quantum chemistry, and quantum machine learning. We first use VAns in the Variational Quantum Eigensolver (VQE) algorithm~\cite{peruzzo2014variational} to obtain the ground state of the Transverse Field Ising model (TFIM), the XXZ Heisenberg spin model, and the  $H_2$ and $H_4$ molecules. We then apply VAns to a quantum autoencoder~\cite{romero2017quantum} task for data compression. We then move to use VAns to compile a Quantum Fourier Transform unitary in systems up to 10 qubits. Finally, we benchmark the performance of VAns under the presence of noisy channels, which are unavoidable in quantum hardware, and demonstrate that it successfully finds cost-minimizing circuits on a wide range of noise-strength levels. 

The simulations presented here were performed using Tensorflow quantum~\cite{broughton2020tensorflow}. Adam~\cite{kingma2015adam} and qFactor~\cite{qFactor} were employed to optimize the continuous parameters $\thv$. While the results shown in noiseless scenarios were obtained from \textit{a single instance} of the algorithm for each problem, noisy simulations required several instances of the algorithm to reach a minima (here we present results obtained after 50 iterations), due to the fact that the optimization landscape is considerably more challenging in the latter case. The dictionary of gates used consisted on single-qubit rotation around $x$ and $z$ axis, and CNOTs gates between all qubits in the circuit. Moreover, we have assumed no connectivity constraints under the quantum circuits under consideration. In the following examples, VAns was initialized to either a separable ansatz or an $\ell$-layer HEA, with $\ell<3$ (see Fig.~\ref{fig:initialansatz} for a depiction of separable and 2-layer HEA circuits); in our heuristics we observed that, as expected,  varying the initial circuit helps to attain a quicker convergence towards optimal cost-values. Nevertheless, the \textit{optimal} choice of initial circuit is subtle: highly expressible initial circuits are not always the best choice, since they might bias the search and even diminish the convergence rate due to the appearance of multiple local minima in the optimization landscape. Moreover, under the presence of noise, long-depth initial circuits such as HEA potentially increase the cost-value function instead of diminishing it, due to noise accumulation. In such a case, it takes VAns a higher amount of iterations to shorten circuit's depth. For the noisy simulations we have considered a one-layered HEA as initial circuit for VAns.

\subsection{Transverse Field Ising model}
Let us now consider a cyclic TFIM chain. The Hamiltonian of the system reads
\begin{equation}\label{eq:HTFIM}
    H=-J\sum_{j=1}^n \sigma_j^x\sigma_{j+1}^x-g\sum_{j=1}^n \sigma_j^z\,,
\end{equation}
where $\sigma_j^{x(z)}$ is the Pauli  $x$ ($z$) operator acting on qubit $j$, and where $n+1\equiv 1$ to indicate periodic boundary conditions. Here, $J$ indicates the interaction strength, while $g$ is the magnitude of the transverse magnetic field. As mentioned in Section~\ref{sec:rw}, when using the VQE algorithm the goal is to optimize a parametrized quantum circuit $U(\kvec,\thv)$ to prepare the ground state of $H$ so that the cost function becomes $E(\kvec,\vec{\theta})=\Tr[H U(\kvec,\thv)\rho U\ad (\kvec,\thv)]$, where one usually employs $\rho=\dya{\vec{0}}$ with $\ket{\vec{0}}=\ket{0}^{\otimes n}$. Note that we here employ $E$ as the cost function label to keep with usual notation convention.

\begin{figure}[t!]
\centering
\includegraphics[width=1\columnwidth]{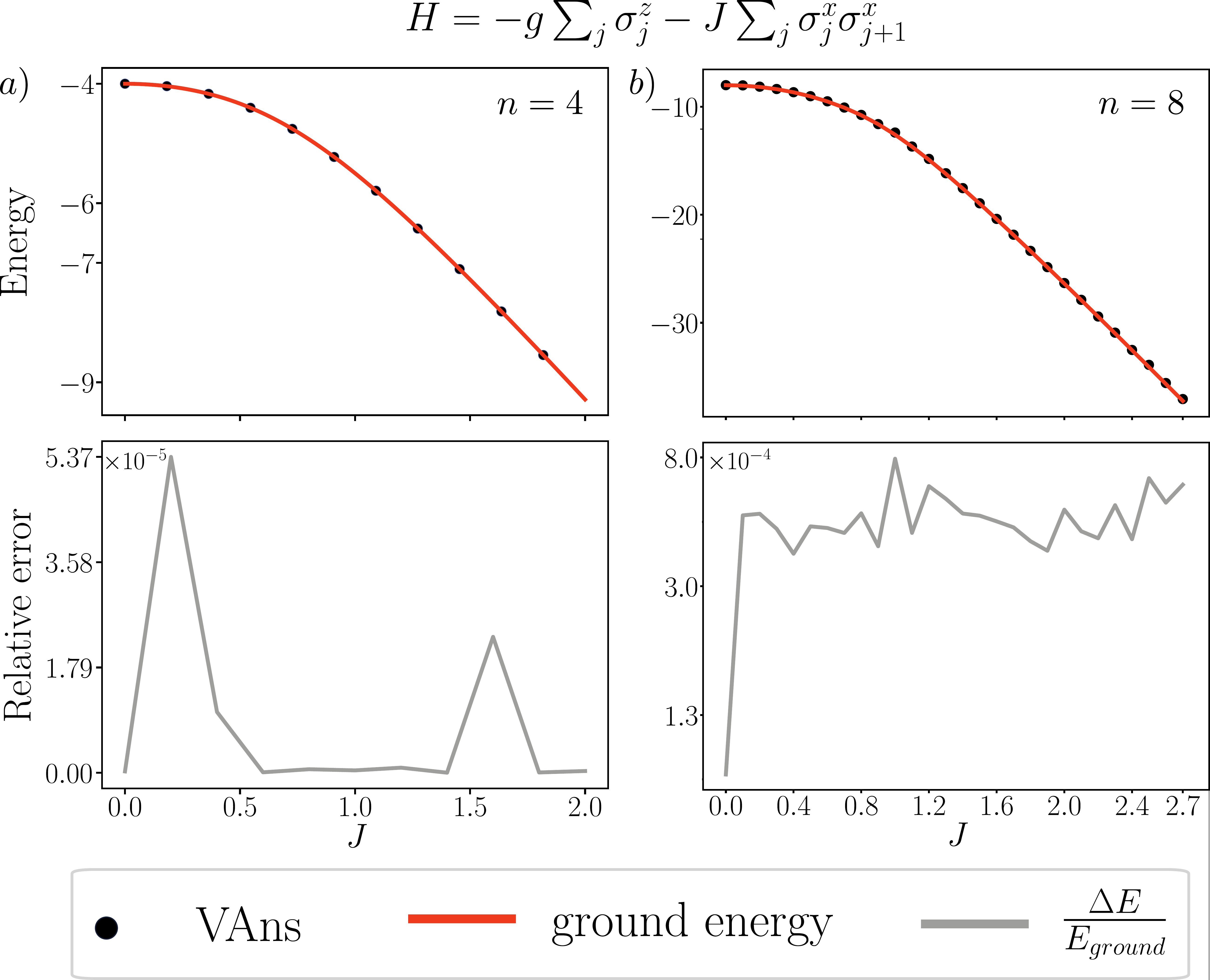}
\caption{\textbf{Results of using VAns to obtain the ground state of a Transverse Field Ising model}. Here we use VAns in the VQE algorithm for the Hamiltonian in~\eqref{eq:HTFIM} with (a) $n=4$ qubits and (b) $n=8$ qubits, field $g=1$, for different values of the interaction $J$. Top panels: solid lines indicate the exact ground state energy, and the markers are the energies obtained using VAns. Bottom panels: Relative error in the energy for the same interaction values.  }
\label{fig:TFIM}
\end{figure}

In Fig.~\ref{fig:TFIM} we show results obtained from employing VAns to find the ground state of a TFIM model of Eq.~\eqref{eq:HTFIM} with $n=4$ qubits (a) and with $n=8$ qubits (b), field $g=1$, and different interactions values. To quantify the performance of the algorithm, we  additionally show the relative error $\left|\Delta E/E_0\right|$, where $E_0$ is the exact ground state energy $E_0$, $\Delta E=E_{\text{VAns}}-E_0$, and $E_{\text{VAns}}$ the best energy obtained through VAns. For $4$ qubits, we see from Fig.~\ref{fig:TFIM} that the relative error is always smaller than $6\times 10^{-5}$, showing the that ground state was obtained for all cost values. Then, for $n=8$ qubits, VAns obtains the ground state of the TFIM with relative error smaller than $8\times 10^{-4}$. 

\begin{figure}[h!]
\centering
\includegraphics[width=.9\columnwidth]{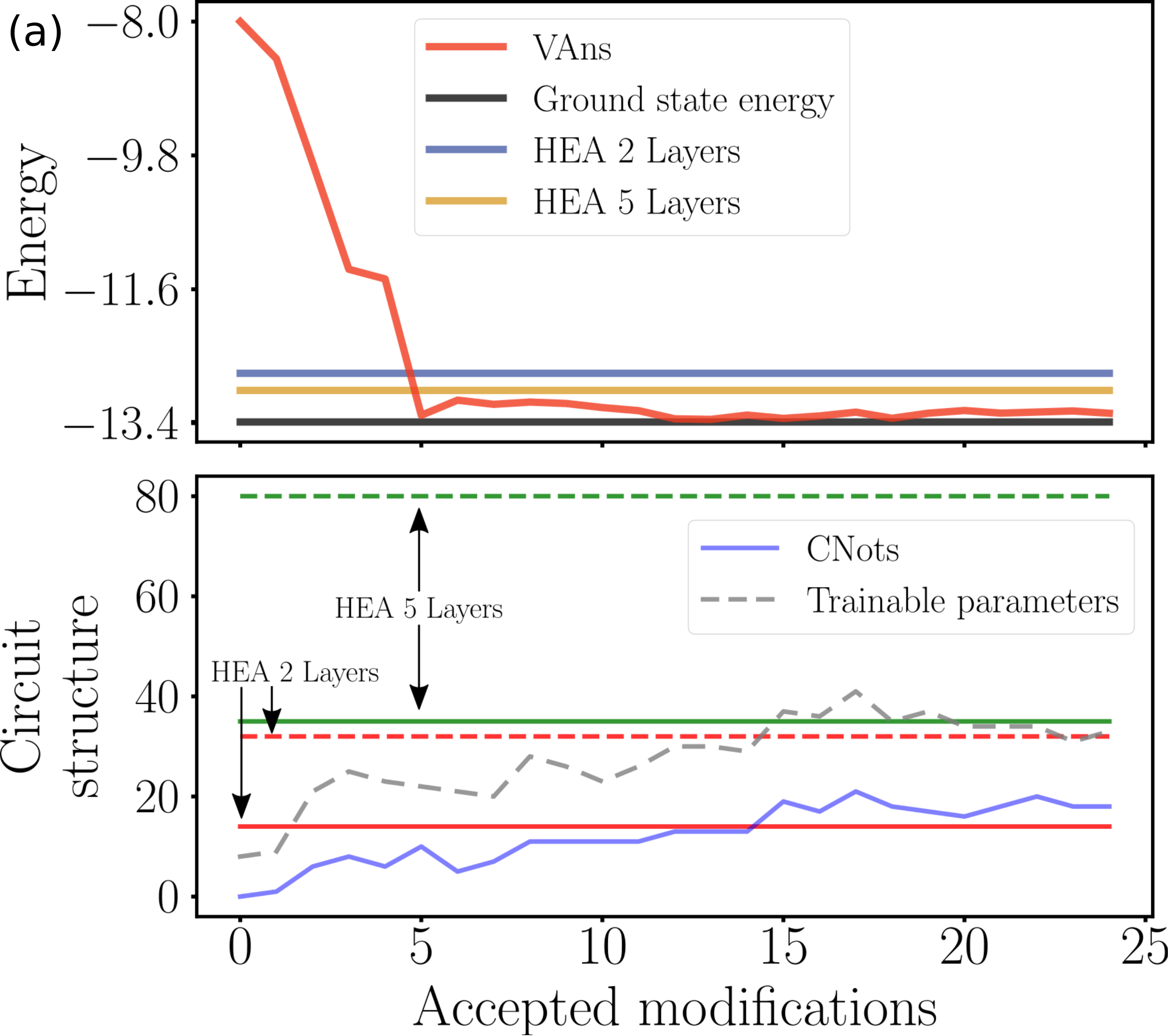}
\includegraphics[width=.9\columnwidth]{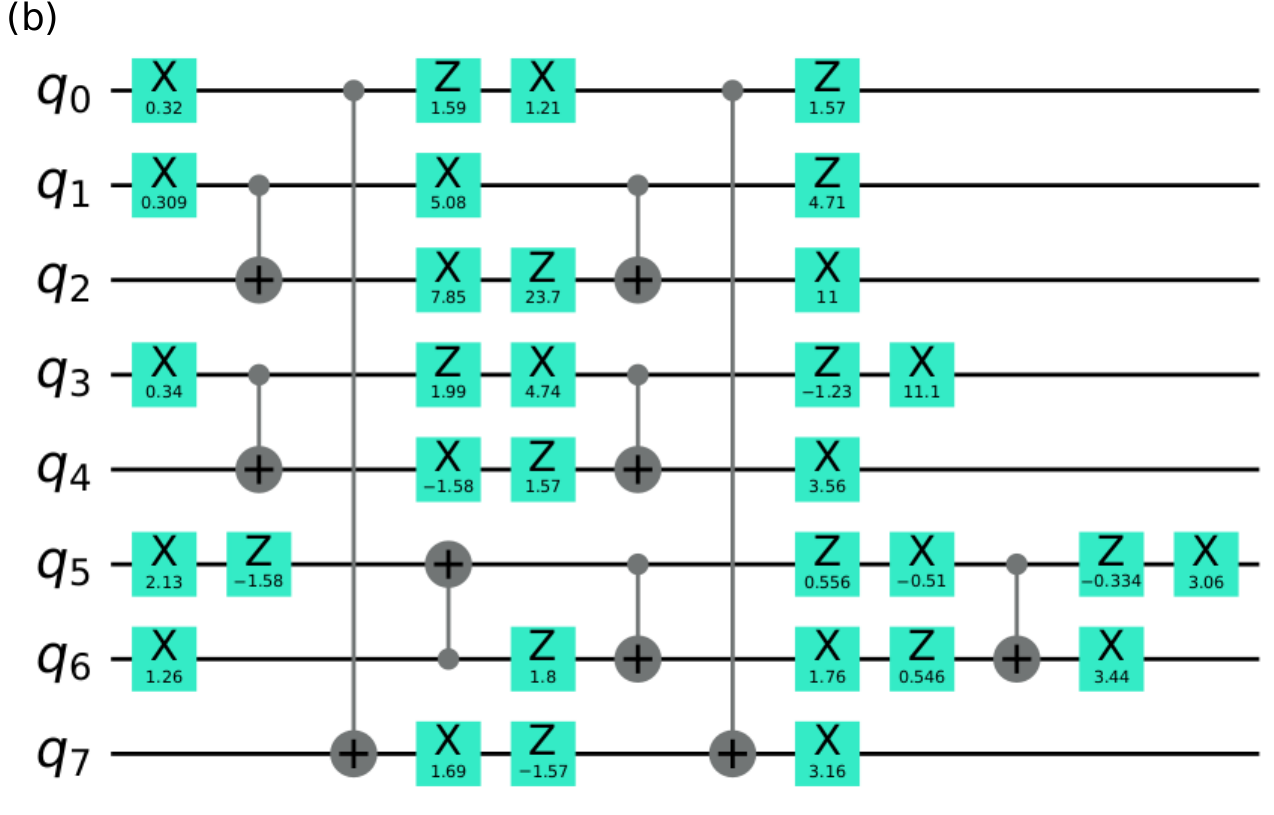}
\caption{\textbf{VAns learning process}. \textit{(a)} Here we show an instance of running the algorithm for the Hamiltonian in Eq.~\eqref{eq:HTFIM} with $n=8$ qubits, field $g=1$, and interaction $J=1.5$. The top panel shows the cost function value and the bottom panel depicts the number of CNOTs, and the number of trainable parameters versus the number of modifications of the ansatz accepted in the VAns algorithm. Top: As the number of iterations increases, VAns minimizes the energy until one finds the ground state of the TFIM. Here we also show the best results obtained by training a fixed structure layered Hardware Efficient Ansatz (HEA) with $2$ and $5$ layers, and in both cases, VAns outperforms the HEA. Bottom: While initially the number of CNOTs and number of trainable parameters increases, the \texttt{Simplification} method in VAns prevents the circuit from constantly growing, and can even lead to shorter depth circuits that achieve better solutions. Here we also show the number of CNOTs (solid line) and parameters (dashed line) in the HEA ansatzes considered, and we see that VAns can obtain circuits with less entangling and trainable gates. \textit{(b)} We show a low-depth, ground-state preparing circuits found by VAns during the learning process; here $Z$ ($X$) indicates a rotation about the $z$ ($x$) axis, about the corresponding value appearing below. }
\label{fig:learning}
\end{figure}

To gain some insight into the learning process, in Fig.~\ref{fig:learning} we show the cost function value, number of CNOTs, and the number of trainable parameters in the circuit discovered by VAns as different modifications of the ansatz are accepted to minimize the cost in an $n=8$ TFIM VQE implementation. Specifically, in Fig.~\ref{fig:learning}(top) we see that as VAns explores the architecture hyperspace, the cost function value continually decreases until one can determine the ground state of the TFIM. Fig.~\ref{fig:learning}(bottom) shows that initially VAns increases the number of trainable parameters and CNOTs in the circuit via the \texttt{Insertion} step. However, as the circuit size increases, the action of the \texttt{Simplification} module becomes more relevant as we see that the number of trainable parameters and CNOTs can decrease throughout the computation. Moreover, here we additionally see that reducing the number of CNOTs and trainable parameters can lead to improvements in the cost function value. The latter indicates that VAns can indeed lead to short depth ansatzes which can efficiently solve the task at hand, even without the presence of noise.

In Fig.~\ref{fig:learning} we also compare the performance of VAns with that of the layered Hardware Efficient Ansatz of Fig.~\ref{fig:initialansatz}(b) with $2$ and $5$ layers. We specifically compare against those two fixed structure ansatzes as the first (latter) has a number of trainable parameters (CNOTs) comparable to those obtained in the VAns circuit. In all cases, we see that VAns can produce better results than those obtained with the Hardware Efficient Ansatz.

\subsection{$XXZ$ Heisenberg Model}
\begin{figure}[h!]
\centering
\includegraphics[width=1\columnwidth]{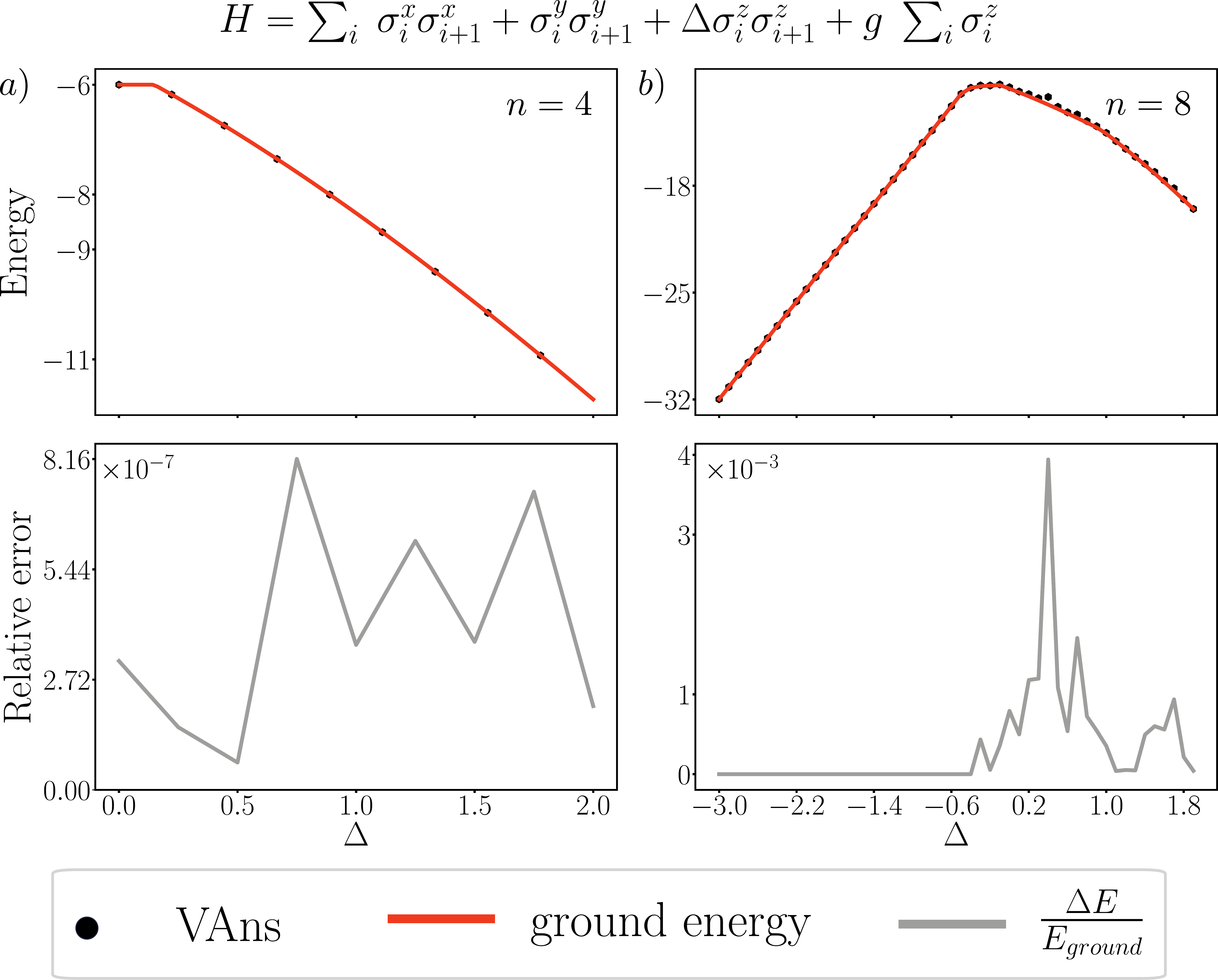}
\caption{\textbf{Results of using VAns to obtain the ground state of a Heisenberg $XXZ$ model}. Here we use VAns in the VQE algorithm for the Hamiltonian in~\eqref{eq:HXXZ} with (a) $n=4$ and (b) $n=8$ qubits, field $g=1$, and indicated anisotropies $\Delta$. Top panels: The solid line indicates the exact ground state energy, and the markers are the energies obtained using VAns. Bottom panels: Relative error in the energy for the same anisotropy values.     }
\label{fig:XXZ}
\end{figure}

Here we use VAns in a VQE implementation to obtain the ground state of a periodic $XXZ$ Heisenberg spin chain in a transverse field. The Hamiltonian of the system is
\begin{equation}\label{eq:HXXZ}
    H=\sum_{j=1}^n \sigma_j^x\sigma_{j+1}^x+\sigma_j^y\sigma_{j+1}^y+\Delta \sigma_j^z\sigma_{j+1}^z+g\sum_{j=1}^n \sigma_j^z\,,
\end{equation}
where again $\sigma_j^{\mu}$ are the Pauli operator (with $\mu=x,y,z$) acting on qubit $j$, $n+1\equiv 1$ to indicate periodic boundary conditions, and where $\Delta$ is the anisotropy.  We recall that $H$ commutes with the total spin component $S_z=\sum_j \sigma^z_i$, meaning that its eigenvectors have definite magnetization $M_Z$ along $z$~\cite{cerezo2017factorization}.

In Fig.~\ref{fig:XXZ} we show numerical results for finding the ground state of~\eqref{eq:HXXZ} with $n=4$ and $n=8$ qubits, field $g=1$, and for different anisotropy values. For $4$ qubits, we see that VAns can obtain the ground state with relative errors which are always smaller than $9\times 10^{-7}$. In the $n=8$ qubits case, the relative error is of the order $10^{-3}$, with error increasing in the region $0<\Delta<1$. We remark that a similar phenomenon is observed in~\cite{cervera2020meta}, where errors in preparing the ground state of the $XXZ$ chain increase in the same region. The reason behind this phenomenon is that the optimizer can get stuck in a local minimum where it prepares excited states instead of the ground state. Moreover, it can be verified that while the ground state and the three first exited states all belong to the same magnetization sub-space of state with magnetization $M_Z=0$, they have in fact different local symmetries and structure. Several of the low-lying excited states have a N\'eel-type structure of spins with non-zero local magnetization along $z$ of the form $|\uparrow\downarrow\uparrow\downarrow\cdots\rangle$. On the other hand, the state that becomes the ground-state for $\Delta>1$ is a state where all spins have zero local magnetization along $z$, meaning that the local states are in the $xy$ plane of the Bloch sphere. Since there is a larger number of excited states with a N\'eel-type structure (and with different translation symmetry) variational algorithms tend to prepare such states when minimizing the energy. Moreover, since mapping a state  with non-zero local magnetization along $z$ to a state with zero local magnetization requires a transformation acting on all qubits, any algorithm performing local updates will have a difficult time finding such mapping.

\subsection{Molecular Hamiltonians}

Here we show results for using VAns to obtain the ground state of the Hydrogen molecule and the $H_4$ chain. Molecular electronic Hamiltonians for quantum chemistry are usually obtained in the second quantization formalism in the form
\begin{equation}\label{eq:Hferm}
    H=\sum_{mn} h_{mn} a_m\ad a_n +\frac{1}{2}\sum_{mnpq} h_{mnpq} a_m\ad a_n\ad a_p a_q\,, 
\end{equation}
where $\{a_m\ad\}$ and $\{a_n\}$ are the fermionic creation and annihilation operators, respectively, and where the coefficients $h_{mn}$ and $h_{mnpq}$ are one- and two-electron overlap integrals, usually computed through classical simulation methods~\cite{aspuru2005simulated}. To implement~\eqref{eq:Hferm} in a quantum computer one needs to map the fermionic operators into qubits operators (usually through a Jordan Wigner or  Bravyi-Kitaev transformation). Here we employed the OpenFermion package~\cite{mcclean2019openfermion} to map~\eqref{eq:Hferm} into a Hamiltonian expressed as a linear combination of $n$-qubit Pauli operators of the form
\begin{equation}\label{eq:HfermJW}
    H=\sum_{\vec{z}} c_{\vec{z}} \sigma^{\vec{z}}\,, 
\end{equation}
with $\sigma^{\vec{z}}\in\{\id,\sigma^x,\sigma^y,\sigma^z\}^{\otimes n}$, $c_{\vec{z}}$ real coefficients, and $\vec{z}\in\{0,x,y,z\}^{\otimes n}$. 

In all cases, the basis set used to approximate atomic orbitals was STO-3g and a neutral molecule was considered. The Jordan-Wigner transformation was used in all cases. While for the $H_2$ the number of qubits required is four ($n=4$), this number is doubled for the $H_4$ chain ($n=8$).

\subsubsection{$H_2$ Molecule}

%\subsection{Hydrogen Molecule}

\begin{figure}[t!]
\centering
\includegraphics[width=.9\columnwidth]{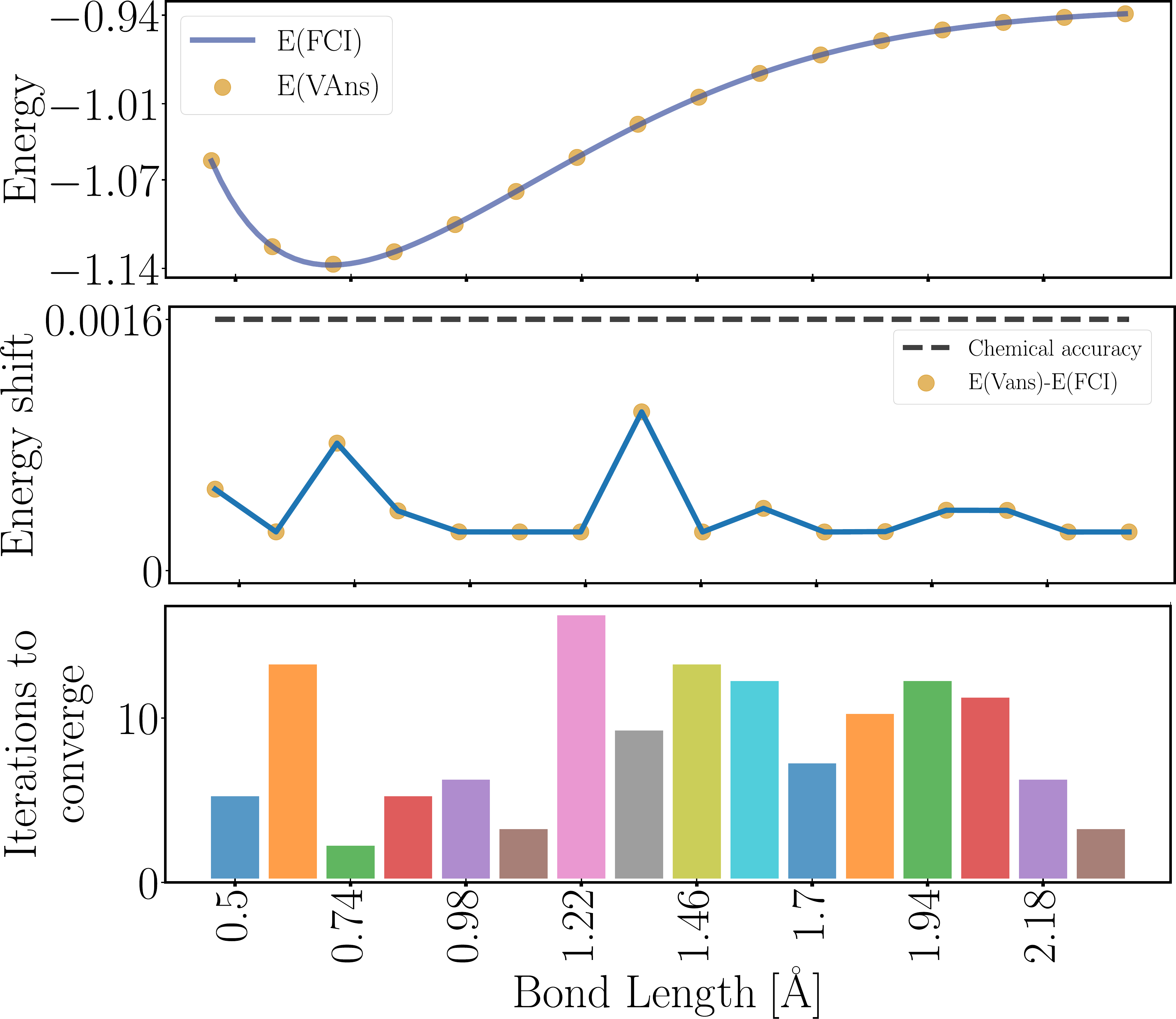}
\caption{\textbf{Results of using VAns to obtain the ground state of a Hydrogen molecule, at different bond lengths}. Here we use VAns in the VQE algorithm for the molecular Hamiltonian obtained after a Jordan-Wigner transformation, leading to a 4-qubit circuit. Top: Solid lines correspond to ground state energy as computed by the Full Configuration Interaction (FCI) method, whereas points correspond to energies obtained using VAns. Middle: Differences between exact and VAns ground state energies are shown. Dashed line corresponds to chemical accuracy, which stands for the ultimate accuracy experimentally reachable in such systems. Bottom: Number of iterations required by VAns until convergence are shown.}
\label{fig:H2}
\end{figure}

Figure~\ref{fig:H2} shows the results obtained for finding the ground state of the Hydrogen molecule at different bond lengths. As shown, VAns is always able to accurately prepare the ground state within chemical accuracy. Moreover, as seen in Figure~\ref{fig:H2}(bottom), VAns usually requires less than $15$ iterations until achieving convergence, showing that the algorithm quickly finds a way through the architecture hyperspace towards a solution.

\subsubsection{$H_4$ Molecule}

\begin{figure}[t!]
\centering
\includegraphics[width=.9\columnwidth]{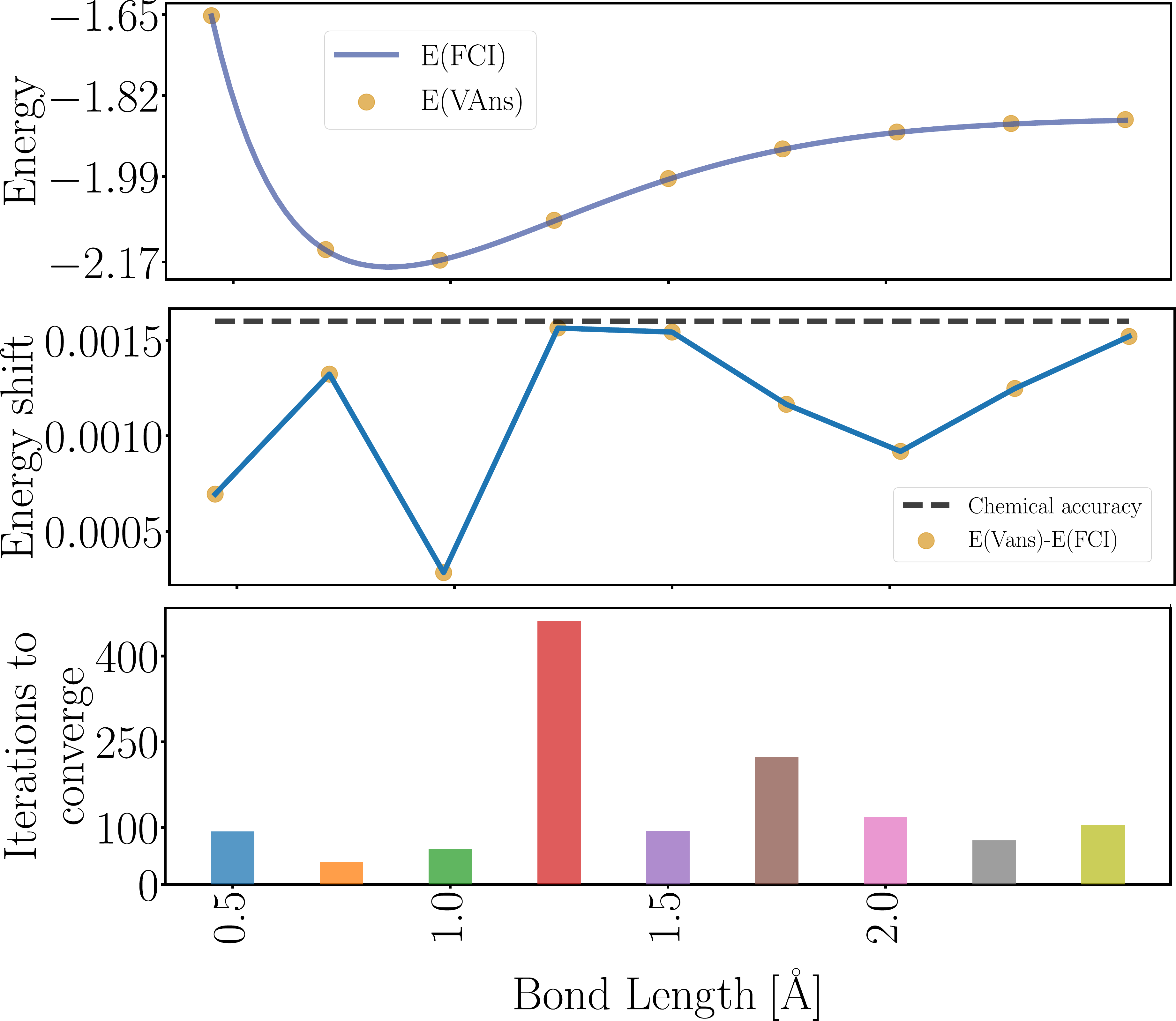}
\caption{\textbf{Results of using VAns to obtain the ground state of a $H_4$ chain with a linear geometry, at different bond lengths}. Here we use VAns in the VQE algorithm for the molecular Hamiltonian obtained after a Jordan-Wigner transformation, leading to an 8-qubit circuit. Top: Solid lines correspond to ground state energy as computed by the Full Configuration Interaction (FCI) method, whereas points correspond to energies obtained using VAns. Middle: Differences between exact and VAns ground state energies are shown. Dashed line corresponds to chemical accuracy. Bottom: Number of iterations required by VAns until convergence are shown.}
\label{fig:H4}
\end{figure}

Figure~\ref{fig:H4} shows the results obtained for finding the ground state of the $H_4$ chain, where equal bond distances are taken. Noticeably, the dictionary of gates $\mathcal{D}$ chosen here is not a \textit{chemical-inspired} one (e.g., it does not contain single and double excitation operators nor its hardware-efficient implementations), yet VAns is able to find ground-state preparing circuits within chemical accuracy~\cite{mcardle2020quantum}. 

\subsection{Quantum Autoencoder}

\begin{figure*}[t!]
\centering
\includegraphics[width=1.\linewidth]{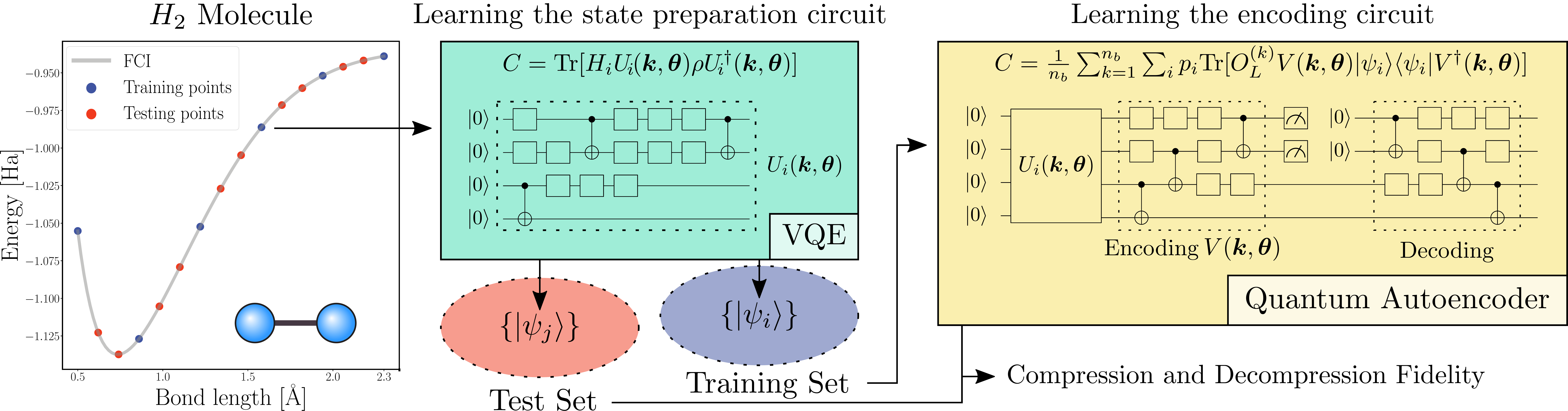}
\caption{\textbf{Schematic diagram of the quantum autoencoder implementation}. We first employ VAns to learn the circuits that prepare the ground states $\{\ket{\psi_i}\}$ of the $H_2$ molecule for different bond lengths. These ground states are then used to create a training set and test set for the quantum autoencoder implementation. The goal of the autoencoder is to train an encoding parametrized quantum circuit $V(\kvec, \thv)$ to compress each $\ket{\psi_i}$ into a subsystem of two qubits so that one can recover $\ket{\psi_i}$ from the reduced states. The performance of the autoencoder can be quantified by computing the compression and decompression fidelities, i.e., the fidelity between the input state and the output state to the encoding/decoding circuit.     }
\label{fig:AE}
\end{figure*}
 
Here we discuss how to employ VAns and the results from the previous section for the Hydrogen molecule to train the quantum autoencoder introduced in~\cite{romero2017quantum}. For the sake of completeness, we now describe the quantum autoencoder algorithm for compression of quantum data.

Consider a bipartite quantum system $AB$ of $n_A$ and $n_B$ qubits, respectively.  Then, let $\{p_i,\ket{\psi_i}\}$ be a training set of pure states on $AB$. The goal of the quantum autoencoder is to train an encoding parametrized quantum circuit $V(\kvec, \thv)$ to compress the  states in the training set onto subsystem $A$, so that one can discard the qubits in subsystem $B$ without losing information. Then, using the decoding circuit (simply given by $V\ad(\kvec, \thv)$) one can recover each state $\ket{\psi_i}$ in the training set (or in a testing set) from the information in subsystem $A$.  Here, $V(\kvec, \thv)$ is essentially decoupling subsystem $A$ from subsystem $B$, so that the state is completely compressed into subsystem $A$ if the qubits in $B$ are found in a fixed target state (which we here set as $\ket{\vec{0}}_B=\ket{0}^{\otimes n_B}$). 

As shown in~\cite{romero2017quantum}, the degree of compression can be quantified by the cost function
\begin{equation}
    C_G(\kvec,\thv)=1-\sum_ip_i\Tr\left[\left(\dya{\vec{0}}_B\otimes \id_A\right)V\dya{\psi_i}V\ad\right]\,,\nonumber
\end{equation}
where $\id_B$ is the identity on subsystem $A$, and where we have omitted the $\kvec$ and $\thv$ dependence in $V$ for simplicity of notation. Here we see that if the reduced state in $B$ is $\ket{\vec{0}}_B$ for all the states in the training set, then the cost is zero. Note that, as proved in~\cite{cerezo2020cost}, this is a global cost function (as one measures all the qubits in $B$ simultaneously) and hence can have barren plateaus for large problem sizes. This issue can be avoided by considering the following local cost function where one instead measures individually each  qubit in $B$~\cite{cerezo2020cost} 
\small
\begin{equation}\label{eq:local}
    C_L(\kvec,\thv)=1-\sum_{k=1}^{n_B}\sum_i \frac{p_i}{n_B}\Tr\left[\left(\dya{\vec{0}}_k \otimes \id_{A,\overline{k}}\right)V\dya{\psi_i}V\ad\right]\,.
\end{equation}
\normalsize
Here $\id_{A,\overline{k}}$ is the identity on all qubits in $A$ and all qubits in $B$ except for qubit $k$. We remark that this cost function is faithful to $C_G(\kvec,\thv)$,  meaning that both cost functions vanish under the same conditions~\cite{cerezo2020cost}.

As shown in Fig.~\ref{fig:AE}, we employ the ground states $\ket{\psi_i}$ of the $H_2$ molecule (for different bond lengths) to create a training set of six states and a test set of ten states. Here, the circuits obtained through VAns in the previous section serve as (fixed) state preparation circuits for the ground states  of the $H_2$ molecule. We then use VAns to learn an encoding circuit $V(\kvec, \thv)$ which can compress the states $\ket{\psi_i}$ into a subsystem of two qubits. 

\begin{figure}[h!]
\centering
\includegraphics[width=.8\columnwidth]{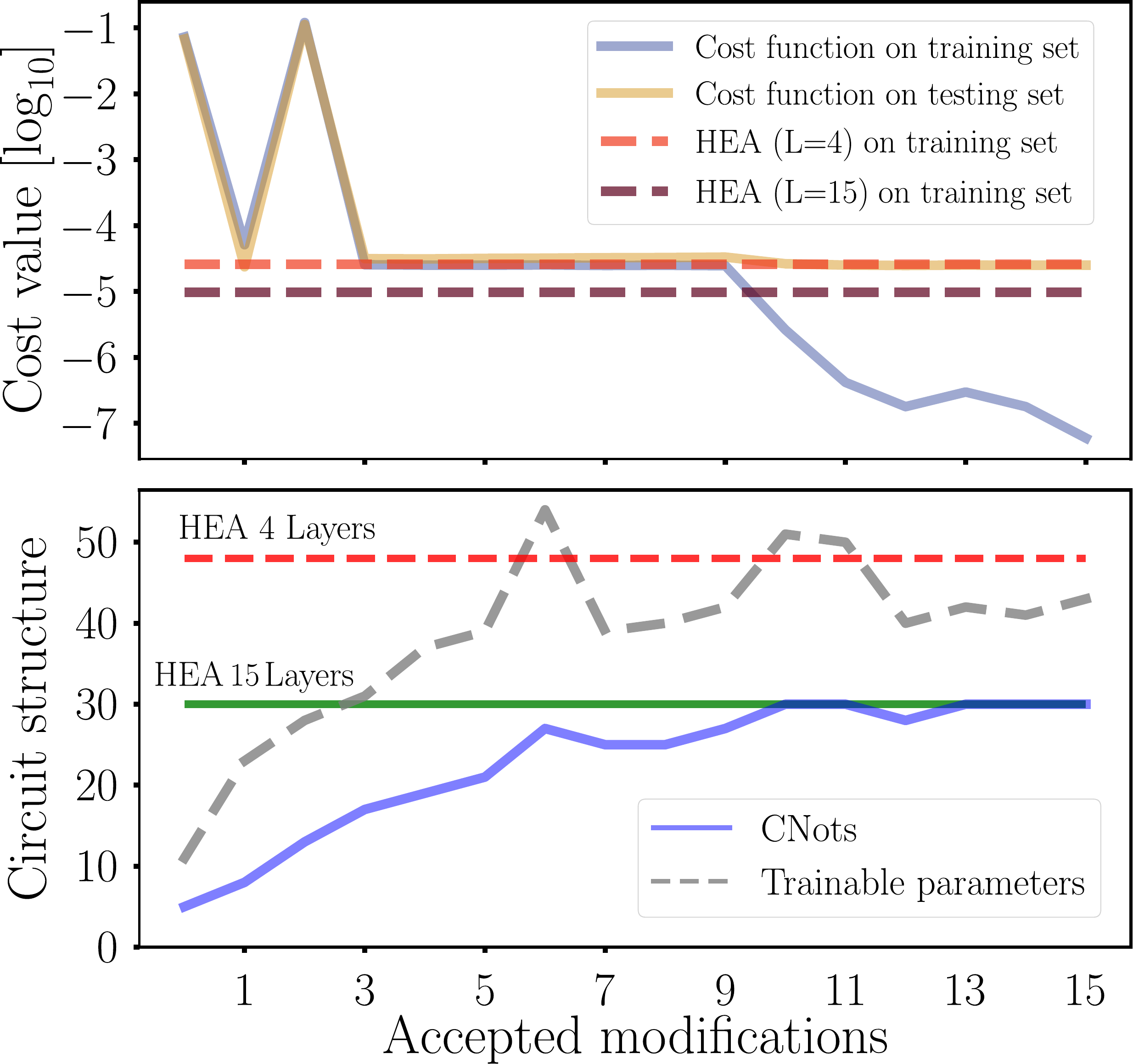}
\caption{\textbf{Results of using VAns to train a quantum autoencoder}. Here we use VAns to train an encoding parametrized quantum circuit by minimizing Eq.~\eqref{eq:local} on a training set comprised of six ground states of the hydrogen molecule. We here also show the lowest cost function obtained for a HEA of $4$ and $15$ layers. Top panel: the cost function evaluated at both versus accepted VAns circuit modifications. In addition, we also show results of evaluating the cost on the testing set. Bottom panel: number of CNOTs, and number of trainable parameters versus the number of modifications of the ansatz accepted in the VAns algorithm. Here we additionally  show the number of CNOTs (solid line) and parameters (dashed line) in the HEA ansatzes considered. We remark that for $L=15$, the HEA has $180$ parameters, and hence the curve is not shown as it would be off the scale. 
}
\label{fig:AE_results}
\end{figure}

Figure~\ref{fig:AE_results} presents results obtained by minimizing the local cost function of~\eqref{eq:local} for a single run of the VAns algorithm. As seen, within $15$ accepted architecture modifications, VAns can decrease the training cost function down to $10^{-7}$, by departing from a separable product ansatz (see Fig.~\ref{fig:initialansatz}). We here additionally show results obtained by training the Hardware Efficient Ansatz of Fig.~\ref{fig:initialansatz}(b) with $4$ and $15$ layers (as they have a comparable number of trainable parameters and CNOTs, respectively, compared to those obtained with VAns). In all cases, VAns achieves the best performance. In particular, it is worth noting that VAns has much fewer parameters ($\sim 45$ versus 180) than the 15-layer HEA, while also achieving a cost value that is lower by two orders of magnitude. Hence, VAns obtains better performance with fewer quantum resources.

Moreover, the fidelities obtained after the decoding process at the training set are reported in Table~\ref{Tab:fi}. Here, $\mathcal{F}$ is used to denote the fidelity between the input state and the state obtained after the encoding/decoding circuit. As one can see, VAns obtains very high fidelities for this task.

\begin{table}[H]
\centering
\begin{tabular}{ |c | c| }
\hline
  Set & $-\log_{10} (1-\mathcal{F})$ \\ 
 \hline
 Training & 6.49 (5.05-6.95) \\  
 \hline
 Testing & 6.17 (3.61-6.95) \\
 \hline
\end{tabular}
\caption{Fidelities for VAns applied to the quantum autoencoder task. The fidelities ($\mathcal{F}$) are reported as Mean value (Min-Max) across both training and testing sets.}
\label{Tab:fi}
\end{table}

\subsection{Unitary compilation} \label{sec:unitary_compilation}

Unitary compilation is a task in which a target unitary is decomposed into a sequence of quantum gates that can be implemented on a given quantum computer. Current quantum computers are limited by the depth of quantum circuits that can be executed on them, which makes the compilation task very important in the near term. Indeed, one wants to decompose a given unitary using as few gates as possible to maximally reduce the effect of noise. In this section we show that VAns is capable of finding very short decompositions as compared with other techniques.

We will illustrate our approach by compiling Quantum Fourier Transform (QFT) on systems up to $n=10$ qubits. Apart from VAns, we also compile the QFT unitary using standard HEA and compare the performance of both methods.

The cost function for unitary compilation is defined as follows. First, a training set is selected
\begin{equation} \label{eq:UC_training_set}
	\{ \big( 
	\ket{\psi_j} , U_\mathrm{QFT}^{(n)} \ket{\psi_j} 
	\big) \}_{j=1}^N \ ,
\end{equation}
where $U_\mathrm{QFT}^{(n)}$ is a target QFT unitary on $n$ qubits and $\ket{\psi_j}$ are $N$, randomly selected input states. We assume that the states $\ket{\psi_j}$ are pairwise orthogonal to avoid potential optimization problems caused by similarities in the training set. The cost function takes the form
\begin{equation} \label{eq:UC_cost}
	C(\kvec,\vec{\theta}) = \sum_{j=1}^N 
	|| U_\mathrm{QFT}^{(n)}\ket{\psi_j} - 
	V(\kvec,\vec{\theta}) \ket{\psi_j} ||^2 \ .
\end{equation}
Note that the cost function introduced in Eq.~\eqref{eq:UC_cost} becomes equivalent to a more standard one, $C'(\kvec,\vec{\theta}) = || U_\mathrm{QFT}^{(n)} - V(\kvec,\vec{\theta})||^2$, when $N=2^n$. While $C(\kvec,\vec{\theta})$ measures the distance between the exact output of QFT and the one returned by $V(\kvec,\vec{\theta})$ only on selected input states, $C'(\kvec,\vec{\theta})$ measures the discrepancy between full unitaries $U_\mathrm{QFT}^{(n)}$ and $V(\kvec,\vec{\theta})$.

It has recently been shown~\cite{caro2021generalization} that $N \ll 2^n$ is sufficient to accurately decompose $U_\mathrm{QFT}^{(n)}$. More precisely, a constant number of training states $N$ (independent of $n$) can be used to ensure small value of $C'(\kvec,\vec{\theta})$, while minimizing the cost function in Eq.~\eqref{eq:UC_cost}.
This observation provides an exponential speedup in evaluating the cost function for unitary compilation. Indeed, the cost of evaluating $C(\kvec,\vec{\theta})$ in Eq.~\eqref{eq:UC_cost} is $N \cdot 2^n$ (assuming the circuit $V(\kvec,\vec{\theta})$ consists of few body gates and the states $U_\mathrm{QFT}^{(n)} \ket{\psi_j} $ are given in computational basis), while the cost of computing $C'(\kvec,\vec{\theta})$ is $4^n$.

The number of training states $N$ which lead to small value of $C'(\kvec,\vec{\theta})$ depends on the number of independent variational parameters in $V(\kvec,\vec{\theta})$. Suboptimal decompositions $V(\kvec,\vec{\theta})$ (in terms of number of parametrized gates) will require larger $N$ to achieve good compilation accuracy. We have used $N = 15$ for $n=10$ qubit compilation with VAns, and a much larger training set in an approach that uses HEA; the increase in $N$ is necessary since the latter approach needed a much deeper circuit, as discussed below.

We have used most general two-qubit gates as the building block in VAns and to construct HEA. This is a slight generalization to the \texttt{Insertion} and \texttt{Simplification} steps in the VAns algorithm discussed above. % The optimization over $\vec{\theta}$ was performed with qFactor~\cite{qFactor}.
General two-qubit gates can be decomposed down to CNOTs and one-qubit rotations using standard methods.

The method based on HEA requires very deep circuits (at $n=10$). They consists of so many gates that the regular optimization has very small success probability. We therefore modify the method based on HEA and utilize the recursive structure of $U_\mathrm{QFT}^{(n)}$. In the modified approach, we use HEA to compile $U_\mathrm{QFT}^{(n-1)}$ and then use it to create an ansatz for $U_\mathrm{QFT}^{(n)}$. The ansatz for larger system size additionally consists of several layers of HEA. We apply the above growth technique starting from $n=3$ to eventually build the ansatz for $n=10$. We stress that VAns does not require such simplification and is capable of finding the decomposition  with high success probability directly at $n=10$ while initialized randomly. 

\begin{figure}[t!]
\centering
\includegraphics[width=\columnwidth]{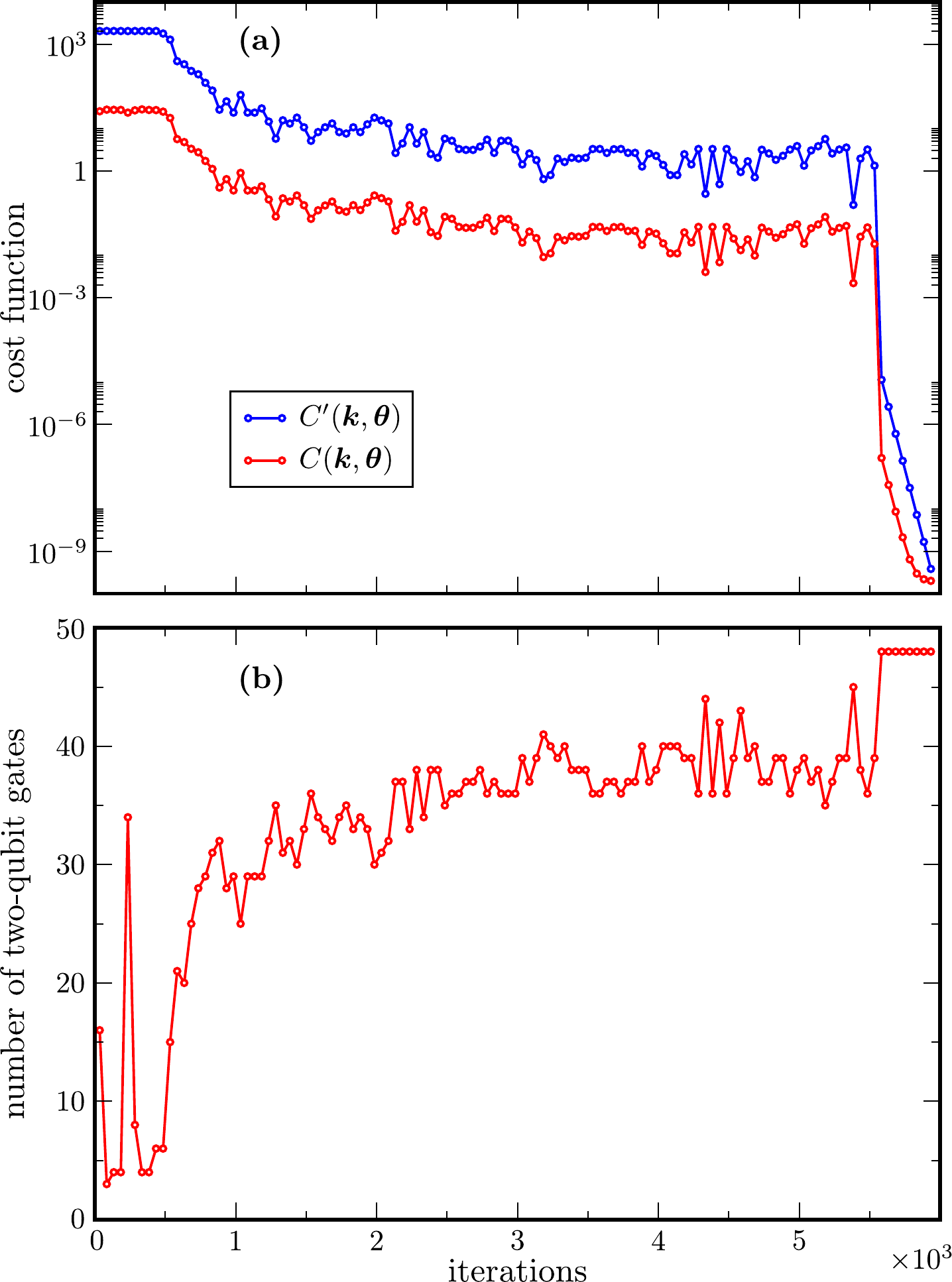}
\caption{\textbf{Results of using VAns for unitary compilation}. %[DRAFT]
Here we use VAns to find a decomposition of QFT unitary defined on $n=10$ qubits, by minimizing a cost function $C(\kvec,\vec{\theta})$ (red line in panel {\bf (a)}) defined in Eq.~\eqref{eq:UC_cost}. The cost evaluates a difference between exact output of QFT and the one returned by a current circuit, on a small number of input states only ($N=15$). The blue line shows corresponding difference between full unitaries, $C'(\kvec,\vec{\theta}) = || U_\mathrm{QFT}^{(n)} - V(\kvec,\vec{\theta})||^2$. We observe a high correlation between those two cost functions. Panel {\bf (b)} shows how VAns modifies the number of two-qubit gates as it approaches the minimum of $C(\kvec,\vec{\theta})$. The minimum is found with 48 gates, which is $\sim 4.5$ times less than the decomposition found with HEA (not shown).
}
\label{fig:unitary_compilation_results}
\end{figure}

Figure~\ref{fig:unitary_compilation_results} shows VAns results for $n=10$ qubit QFT compilation. Panel (a) depicts how the value of the cost function $C(\kvec,\vec{\theta})$ is minimized over the iterations. We also show the corresponding value of $C'(\kvec,\vec{\theta}) = || U_\mathrm{QFT}^{(n)} - V(\kvec,\vec{\theta})||^2$. We observe strong correlation between both cost functions. $C'(\kvec,\vec{\theta})$ is eventually minimized below $10^{-9}$ at the end of the optimization. Panel (b) shows how the number of two-qubit gates evolves as VAns optimization is performed. Excluding the initial ``warm-up'' period, during which the cost function has very large (close to maximal possible) value, the number of two-qubit gates %def 2-qubit gate, are the general 2-qubit gates? in terms of CNOTs? 12 each right?
is steadily grown reaching 48 at the end of the optimization.
%That number is only slightly larger than the number of two-qubit gates (45) used in the textbook QFT circuit for $n=10$.

The approach based on HEA requires 219 general two-qubit gates to decompose $n=10$ QFT, which is over 4 times more than the best circuit found by VAns. The HEA approach uses recursive structure of QFT to find accurate decomposition, while VAns does not rely on that property and finds a solution in fewer number of iterations. Finally, VAns takes advantage of the generalization bound \cite{caro2021generalization}, finding solution with near optimal number of variational parameters in $V(\kvec,\vec{\theta})$; the training set size $N$ required for small generalization error is small resulting in fast cost function evaluations.

\subsection{Noisy simulations}
The results previously shown were obtained without considering hardware noise. We observed that VAns was able to better exploit the quantum resources at hand (i.e. attain a lower cost-function value) as compared to its fix-structure counterpart (e.g. HEA). 

We now consider the case where noisy channels are present in the quantum circuit, an unavoidable situation in current experimental setups, with noise essentially forbidding large-depth quantum circuits to preserve quantum coherence. In the context of VQAs, the overall effect of noise is that of degrading the cost-function value and, if its strength is sufficiently high, then short-depth circuits turn to be favoured even at the cost of expressibility. For instance, increasing the number of layers in HEA ansatz might not reduce the cost function and even increase it, since noise accumulates due to the presence of gates.

There are several sources of noise in quantum computers. For instance, experimental implementation of a quantum gate takes a finite amount of time, which in turn depends on the physical qubit at hand, the latter subjected to thermal relaxation errors. Relaxation and dephasing errors depend, in general, on each particular qubit (i.e. \textit{the qubit label}). The overall effect of the gate implementation is often modeled by a depolarizing quantum channel, followed by phase flips and amplitude damping channels, whose strength depends on the aforementioned parameters (gate implementation time, qubit label), gate type and environment temperature. For instance, an entangling gate such as a CNOT injects considerably more noise to the circuit than a single-qubit rotation. Moreover, state preparation and readout errors should be taken into account. We refer the reader to find further details on noise modeling in Ref.~\cite{Georgopoulos2021Modeling}. We also note that additional sources of noise should ultimately be considered, such as idle noise and cross-talk effects~\cite{larose2022error, Prakash2020Software}. Because the complexity of noise modelling in NISQ devices is particularly high, we here propose a sufficiently simple model that yet captures the essential noise sources.

\textit{The $\lambda$-model}. While a noise-model is ultimately linked to the specific quantum hardware at hand and depends on several factors, we propose a unifying and simplified one that depends on a single parameter. The main motivation behind this is that of benchmarking the performance of different ansatzes in the presence of noise. In more complex scenarios, one should consider specific process matrices obtained from \textit{e.g.} process tomography~\cite{Obrien2004Quantum,Zhou2014Process}, which would here obscure the benchmark and also bias it towards specific quantum hardware. In particular, our model is inspired by Refs~\cite{Blank2020Quantum, Georgopoulos2021Modeling}, which in turn are implemented in the \texttt{aer} noise simulator of IBM, and consists of the following models. State preparation and measurement errors are modeled via bit flip channels acting on each qubit, with strength $\lambda \; 10^{-2}$, happening before the circuit $U(\kvec, \thv)$ and measurements respectively. Noise due to gate implementation is modeled as a depolarizing channel, followed by a phase flip and amplitude damping channels acting on the target qubit right after the gate. In principle, the strength of the channel should depend on the specific qubit, and gate type, but in order to keep the model simple enough we have assumed no noise dependence on the qubit label. Moreover, a two-qubit gate is considerably more noisy than a single-qubit one, which in the $\lambda$-model is reflected by the fact that noise strengths are an order of magnitude higher, in the depolarizing channel, than in single-qubit gates, the latter being $\lambda \; 10^{-5}$. Finally, the strengths of the phase flip and amplitude damping channels are set to $\lambda \; 10^{-3}$. We note that, while not considered here, different situations can easily be incorporated such as qubit connectivity constraints, or differences in qubits' quality (some qubits might be noisier than others). In such cases, we expect VAns to find circuits which automatically balance the trade-offs at hand, \textit{i.e.} minimize the number of gates acting on such noisier qubits.

\begin{figure}[t!]
\centering
\includegraphics[width=\columnwidth]{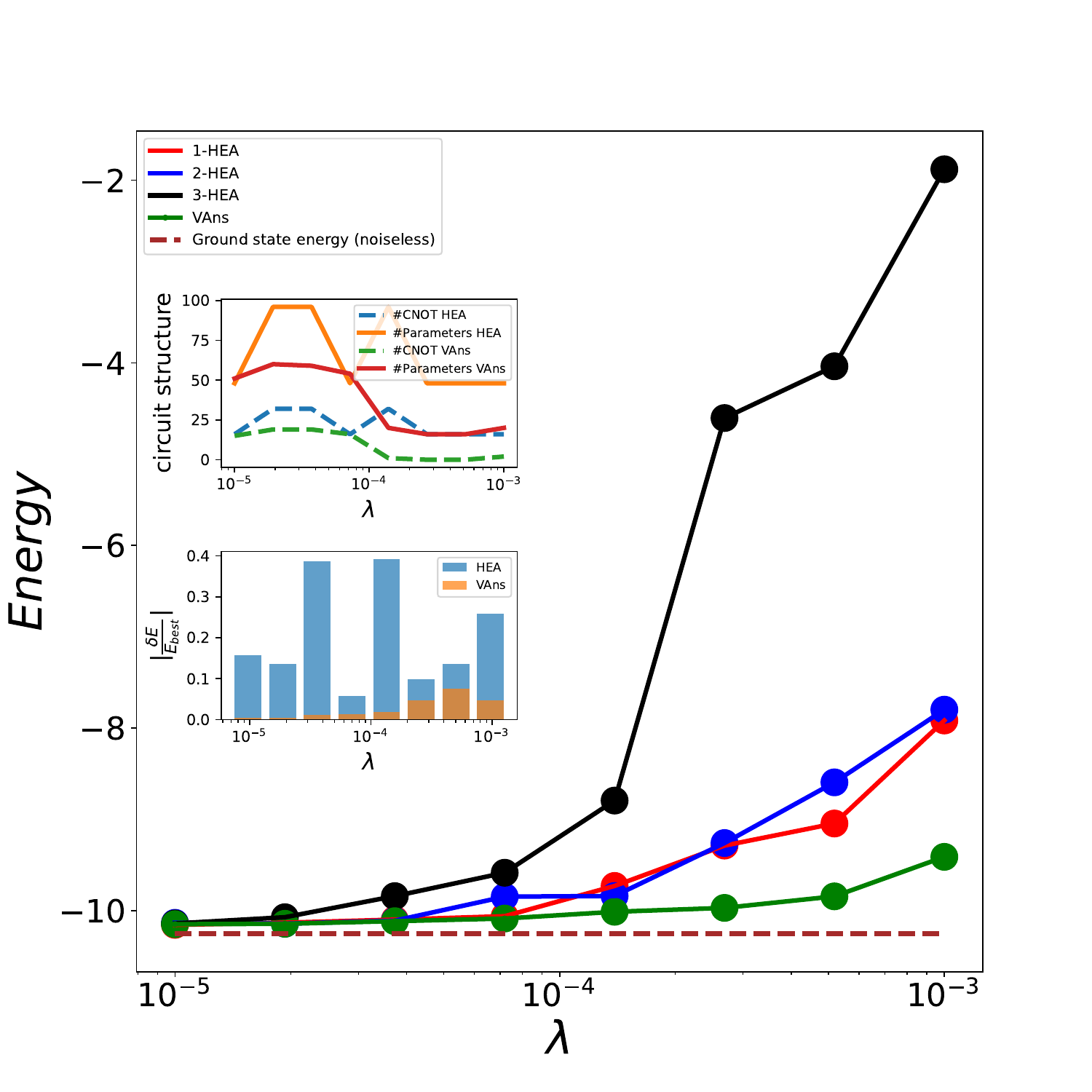}
\caption{\textbf{Results of using VAns for VQE under the $\lambda$-model}. Here we consider the TFIM for 8 qubits, with $g=J=1$. The results are obtained after repeating 50 iterations of optimizations with VAns and HEA respectively. We observe that VAns discovers much more efficient quantum circuits as compared to HEA. As shown in the upper inset, VAns automatically adjusts the circuit layout according to the noise strength at hand, a feature that fix-structure ansatzes lack. In the lower inset we show the relative errors (e.g. standard deviation over optimal cost found) for both ansatzes, across the 50 iterations; we observe that VAns is more precise in reaching a minimum as compared to HEA. We note that in this experiment we have initialized VAns to a 1-layered HEA, which is in turn inconvenient for a sufficiently high value of $\lambda$. Yet, VAns learns how to adapt the ansatz (in this case, finding a separable one) so to reach the lowest cost value. In all cases VAns termination criteria was set to a maximum number of 30 iterations.}
\label{fig:Fig13}
\end{figure}

With this model at hand, we have explored the region of $\lambda$ in which the action of the noise becomes \textit{interesting}. The results of running VAns under the $\lambda$-model for ground state preparation (VQE) of TFIM 8-qubit system are shown in Fig.~\ref{fig:Fig13}. Here, the noise strength is sufficiently high so to affect the ground-state energy (which can otherwise be attained by either VAns or a 3-layered HEA). We thus sweep the value of $\lambda$ by two orders of magnitude, and compare the results that VAns reaches with those of HEA (varying the number of layers of the latter). We see that for a sufficiently high noise strength, increasing the number of gates (e.g. the number of layers in HEA) eventually degrades the cost-function value, as opposed to the noise-less scenario. On the contrary, we observe that even if the noise-strenght is sufficiently high, VAns considerably outperforms HEA by automatically adjusting the depth of the circuit to the noise-strength at hand. Thus, if the noise is large, shallow circuits are found, whereas if the noise-strength is low, deeper circuits are allowed to be explored, since the penalty of adding new gates is smaller. In general, we observe that VAns is capable of finding the best possible circuit under given conditions, which is something that HEA simply can not accomplish. 

\section{Discussion}\label{sec:chau}
In this work we have introduced the VAns algorithm, a semi-agnostic method for building variable structure parametrized quantum circuits. We expect VAns to be especially useful for abstract applications, such as linear systems~\cite{bravo2020variational,huang2019near,xu2019variational}, factoring~\cite{anschuetz2019variational}, compiling~\cite{khatri2019quantum,sharma2019noise}, metrology~\cite{beckey2020variational,koczor2020variational}, and data science~\cite{larose2019variational,cerezo2020variational,biamonte2017quantum,schuld2014quest,abbas2020power,verdon2019quantum}, where physically motivated ansatzes are not readily available. In addition, VAns will likely find use even for physical applications such as finding grounds states of molecular and condensed matter systems, as it provides a shorter depth alternative to physically motivated ansatzes for mitigating the impact of noise, as shown in our noisy simulations.

At each iteration of the optimization, VAns stochastically grows the circuit to explore the architecture hyperspace. More crucially, VAns also compresses and simplifies the circuit by removing redundant gates and unimportant gates.  This is a key aspect of our method, as it differentiates VAns from other variable ansatz alternatives and allows us to produce  short-depth circuits, which can mitigate the effect of noise-induced barren plateaus (NIBPs). We will further investigate this mitigation of NIBPs in future work.

To showcase the performance of VAns, we simulated our algorithm for several paradigmatic problems in VQAs. Namely, we implemented VAns to find ground states of condensed matter systems and molecular Hamiltonians, for a quantum autoencoder problem and for 10-qubit QFT compilation. In all cases, VAns was able to satisfactory create circuits that optimize the cost. Moreover, as expected, these optimal circuits contain a small number of trainable parameters and entangling gates. Here we also compared the result of VAns with results obtained using a Hardware Efficient Ansatz with either the same number of entangling gates, or the same number of parameters, and in all cases, we found that VAns could achieve the best performance. This point is crucial for the success of VAns in the presence of noisy channels, as it automatically adapts the circuit layout to the situation at hand (e.g. noise strength). For instance, under the $\lambda$-model (which is the noise model we have implemented), VAns notably outperforms HEA under ground-state preparation tasks.

While we provided the basic elements and structure of VAns (i.e., the gate \texttt{Insertion}  and gate \texttt{Simplification}  rules), these should be considered as blueprints for variable ansatzes that can be adapted  and tailored to more specific applications. For instance, the gates that VAns inserts can preserve a specific symmetry in the problem. Moreover, one can cast the VAns architecture optimization (e.g., removing unimportant gates) in more advanced learning frameworks. Examples of such frameworks include supervised learning or reinforced learning schemes, which could potentially be employed to detect which gates are the best candidates for being removed. 

%\section{Data availability}\label{sec:data}
%The data supporting the results shown in this paper is available upon reasonable request to the authors. 

\section{ACKNOWLEDGEMENTS}
Some of the simulations presented in this paper were done using Sandbox@Alphabet's TPU-based Floq simulator. MB is thankful to Josep Flix, John Calsamiglia Costa and Carles Acosta for their kindness in facilitating computational resources at Port d’Informació Científica, Barcelona. MB acknowledges support from the U.S. Department of Energy (DOE) through a quantum computing program sponsored by the Los Alamos National Laboratory (LANL) Information Science \& Technology Institute, and support from the Spanish Agencia Estatal de Investigación, project PID2019-107609GB-I00, Generalitat de Catalunya CIRIT 2017-SGR-1127. MC was initially supported by the Laboratory Directed Research and Development (LDRD) program of LANL under project number 20180628ECR, and also supported by the Center for Nonlinear Studies at LANL. PJC acknowledges initial support from the LANL ASC Beyond Moore's Law project. LC was initially supported by the LDRD program of LANL under project number 20200056DR. PJC and LC were also supported by the U.S. Department of Energy (DOE), Office of Science, Office of Advanced Scientific Computing Research, under the Accelerated Research in Quantum Computing (ARQC) program. 

\section{Declarations}
The authors declare no competing interests. All authors contributed equally to this work. Data supporting the claims made in this work is available upon reasonable request.

\bibliography{quantum.bib}

\end{document}